\documentclass[
journal=jpcbfk, 
manuscript=article]{achemso}
\usepackage{amsmath,amssymb}
\usepackage{bm}
\usepackage{graphicx}
\usepackage[T1]{fontenc} 
\usepackage{color}

\usepackage{verbatim}

\author{Soohaeng Yoo Willow}
\affiliation{Department of Chemistry, Illinois Institute of Technology, Chicago, Illinois, 60616}
\author{Bing Xie}
\affiliation{Department of Chemistry, Illinois Institute of Technology, Chicago, Illinois, 60616}
\author{Jason Lawrence}
\affiliation{Department of Computer Science, Illinois Institute of Technology, Chicago, Illinois, 60616}
\author{Robert S. Eisenberg}
\affiliation{Department of Applied Mathematics, Illinois Institute of Technology, Chicago, Illinois, 60616}
\author{David D. L. Minh}
\email{dminh@iit.edu}
\affiliation{Department of Chemistry, Illinois Institute of Technology, Chicago, Illinois, 60616}

\SectionNumbersOn

\title{On the Polarization of Ligands by Proteins}

\begin{document}

\begin{abstract}	
Although ligand-binding sites in many proteins contain a high number density of charged side chains that can polarize small organic molecules and influence binding, the magnitude of this effect has not been studied in many systems. Here, we use a quantum mechanics/molecular mechanics (QM/MM) approach in which the ligand is the QM region to compute the ligand polarization energy of 286 protein-ligand complexes from the PDBBind Core Set (release 2016). We observe that the ligand polarization energy is linearly correlated with the magnitude of the electric field acting on the ligand, the magnitude of the induced dipole moment, and the classical polarization energy. The influence of protein and cation charges on the ligand polarization diminishes with the distance and is below 2 kcal/mol at 9~\AA ~and 1 kcal/mol at 12~\AA. Considering both polarization and solvation appears essential to computing negative binding energies in some crystallographic complexes. Solvation, but not polarization, is essential for achieving moderate correlation with experimental binding free energies.
\end{abstract}

\section{Introduction}

Noncovalent binding to proteins is a key mechanism by which small organic molecules (ligands) interact with biological systems. Most drugs are noncovalent inhibitors of particular targets. Signaling molecules generally bind to specific receptors. Molecules with low solubility often bind to serum albumin. Even in enzymes, noncovalent binding of substrates is a prerequisite to catalysis.

Many proteins generate a strong electrostatic potential that can influence ligand binding. To promote stable folding, globular proteins typically consist of a hydrophobic core and hydrophilic surface. Many amino acids in the latter region are charged. Indeed, in an analysis of 573 enzyme structures, \citet{Jimenez-Morales2012} observed a high number density of oft-charged acidic (aspartic and glutamic acid) and basic (lysine, arginine, and histidine) amino acids in catalytic sites (18.9 $\pm$ 0.58 mol L$^{-1}$) and other surface pockets, including ligand-binding sites (28.2 $\pm$ 0.34 mol L$^{-1}$). For context, the number density of charges is 2.82 $\pm$ 0.03 mol L$^{-1}$ in entire proteins \cite{Jimenez-Morales2012} and 74.3 mol L$^{-1}$ in a sodium chloride salt crystal \cite{haynes2016crc}. Charged amino acid side chains generate patterns in the surrounding electrostatic potential that can have functional roles that include mediating associations with other proteins with complementary electrostatics and channeling charged enzyme substrates \cite{Honig1995}. Within a protein, electrostatic forces can alter redox potentials, shift the pK$_a$s of amino acid residues \cite{Honig1995}, accelerate enzyme catalysis \cite{Garcia-Viloca2003,vanderKamp2007}, and polarize ligands \cite{Hensen2004}.

The importance of ligand polarization in protein-ligand binding has been demonstrated by studies that compare results from similar models with and without polarization. Although the vast majority of current studies modeling biological macromolecules are based on fixed-charge molecular mechanics force fields, polarizable models are being actively developed \cite{Shi2013, Jing2019}.  \citet{Jiao2008} demonstrated that incorporating polarization into a molecular mechanics force field was essential to accurately computing the binding free energy between trypsin and the charged ligands benzamidine and diazamidine. Quantum mechanics (QM) and mixed quantum mechanics/molecular mechanics (QM/MM) methods have also been increasingly employed in predicting the binding pose --- the configuration and orientation of a ligand in a complex --- and binding affinity \cite{Ryde2016, Crespo2017}. Semiempirical QM methods have shown particular promise in correctly distinguishing the native (near-crystallographic) binding pose from decoy poses (non-native poses that have low docking scores) in diverse sets of protein-ligand complexes \cite{Chaskar2014,Pecina2016,Pecina2017,Ajani2017}. QM/MM methods usually couple the QM and MM regions via electrostatic embedding, in which charges from the MM region alter the Hamiltonian in the QM region. Electrostatic embedding allows the QM region (which in most protein-ligand binding studies includes the ligand and sometimes surrounding residues) to polarize in response to charges in the environment. \citet{Cho2005} demonstrated the importance of embedding by evaluating the ability of multiple docking schemes to recapitulate ligand binding poses in 40 diverse complexes. They found that assigning ligand charges using a QM/MM method with electrostatic embedding was generally more successful than a gas-phase QM method without embedding. Subsequently, \citet{Kim2016} performed a more systematic assessment focusing on 40 G protein-coupled receptor crystal structures. The QM/MM method outperformed (1.115 \AA~average RMSD and RMSD<2 \AA~in 36/40 complexes) a gas-phase QM method without embedding (1.672 \AA~average RMSD and RMSD<2 \AA~in 31/40 complexes) and a fixed-charge molecular mechanics method (1.735 \AA~average RMSD and RMSD<2 \AA~in 32/40 complexes). Beyond the context of protein-ligand binding, the inclusion of the polarization energy has been shown to dramatically affect water density \cite{Willow2016} and the structure and dynamics of solvated ions in water clusters \cite{Yoo2003,Chang2006,Caleman2011,Bajaj2016}.

Ligand polarization effects have also been isolated using a decomposition scheme pioneered by \citet{Gao1992}, which was originally applied to the polarization of solutes by aqueous solvents. In this scheme, the polarization energy of molecule $I$, $\Xi_I^\mathrm{pol}$ (Eq. \ref{eq:Epol}), is the sum of the energy from distorting the wave function, $\Xi_I^\mathrm{dist}$ (Eq. \ref{eq:Edist}), and the energy from stabilizing Coulomb interactions relative to the gas phase, $\Xi_I^\mathrm{stab}$ (Eq. \ref{eq:Estab}). For three high-affinity inhibitors of human immunodeficiency virus type 1 (HIV-1) protease, \citet{Hensen2004} found that the magnitude of the ligand polarization energy can be as large as one-third of the electrostatic interaction energy. \citet{Fong2009} considered 6 ligands of HIV-1 protease in near-native poses and found that depending on the level of theory, the polarization energy is from 16\% to 21\% of the electrostatic interaction energy.

Although comparative studies and energy decomposition schemes have strongly indicated the importance of ligand polarization,  the magnitude of this term and the factors contributing to the ligand polarization energy have not, to our knowledge, been investigated for many diverse systems. Here, we address this knowledge gap by calculating the ligand polarization energy for 286 protein-ligand complexes from the PDBBind Core Set (release 2016) \cite{Liu2017c}. The PDBBind is a comprehensive database of complexes for which both Protein Data Bank crystal structures and binding affinity data are available. The Core set is a subset of the PDBBind with high-quality and non-redundant structures meant as a benchmark for molecular docking methods. The size and diversity of this dataset allow us to draw more general and statistically meaningful conclusions about ligand polarization than previous efforts.

\section{Theory and Methods}

\subsection{Energies}

We employed a QM/MM scheme in which the ligand is the QM region and other atoms are the MM region. To enable energy decomposition, the Schr\"{o}dinger equation for the ligand was solved both in the gas phase and with electrostatic embedding.

In the gas phase, the Hamiltonian operator $\hat{H}_I$ of a molecule $I$ is,
\begin{eqnarray}
\label{eq:H_I}
\hat{H}_I = \sum_{i \in I} \frac{1}{2} \frac{\hat{p}_i^2}{m_e} + 
\sum_{i \in I}\sum_{\substack{j > i\\ j \in I}} \frac{1}{r_{ij}} - 
\sum_{i \in I}\sum_{A \in I} \frac{Z_A}{|\pmb{r}_i - \pmb{R}_A|} + 
\sum_{A \in I}\sum_{\substack{B>A\\B \in I}} \frac{Z_A Z_B}{R_{AB}},
\end{eqnarray}
where $i$ and $j$ are indices over all electrons and 
$A$ and $B$ are indices over all atoms in molecule $I$.
$\hat{p}_i$ is the momentum operator and $m_e$ is the mass of an electron.
$\pmb{r}_i$ is the position of electron $i$, 
$\pmb{R}_A$ is the position of atom $A$, and
$Z_A$ is the atomic number of atom $A$.
$r_{ij}$ is the distance between electrons $i$ and $j$, 
and $R_{AB}$ is the distance between atoms $A$ and $B$.
The ground-state energy $E_I$ of the molecule $I$ is
\begin{eqnarray}
E_I = \langle \Psi_I | \hat{H}_I | \Psi_I \rangle, 
\end{eqnarray}
where $\Psi_I$ is the electronic wave function of the molecule $I$.

When the molecule $I$ is placed in an embedding field $Q_I = \{ q_F \}$, the Hamiltonian operator of the embedded molecule is given by $\hat{H}_{I:Q_I} = \hat{H}_I + \hat{H}_{[I/Q_I]}$. We will use $I:Q_I$ to denote the embedding of molecule $I$ in the embedding field $Q_I$. The Hamiltonian operator $\hat{H}_{[I/Q_I]}$ for Coulomb interactions between the molecule $I$ (QM) and the field $Q_I$ (MM) is,
\begin{eqnarray}
\label{eq:H_I_QI}
\hat{H}_{[I/Q_I]} = 
\sum_{i\in I} \sum_{F \in Q_I} \frac{q_F}{|\pmb{r}_i - \pmb{R}_F|} +
\sum_{A \in I}\sum_{F \in Q_I} \frac{Z_A q_F}{|\pmb{R}_A - \pmb{R}_F|},
\end{eqnarray} 
where $F$ is an index over charges in the embedding field. The first summand describes electron-charge interactions and the second proton-charge interactions.
The ground-state energy $E_{I:Q_I}$ of the embedded molecule $I:Q_I$ 
is obtained by
\begin{eqnarray}
E_{I:Q_I} = \langle \Psi_{I:Q_I} | \hat{H}_{I:Q_I} | \Psi_{I:Q_I}\rangle,
\end{eqnarray}
where $\Psi_{I:Q_I}$ is the ground-state electronic wave function of the embedded molecule $I:Q_I$. The embedding field should affect the ground-state wave function of the molecule such that $|\Psi_{I:Q_I}|^2 \neq |\Psi_I|^2$. 

We will use the symbol $\Xi$ to denote a difference between two expectation values. The electronic interaction energy describes the change in electronic energy of a molecule upon interaction with the embedding field,
\begin{equation}
\label{eq:Eelec}
    \Xi_I^\mathrm{elec} = 
        \langle \Psi_{I:Q_I} | \hat{H}_{I:Q_I} | \Psi_{I:Q_I}\rangle - 
        \langle \Psi_I | \hat{H}_I | \Psi_I \rangle.
\end{equation}
\citet{Hensen2004} decomposed $\Xi_I^\mathrm{elec}$ into the polarization energy of a molecule,
\begin{eqnarray}
\label{eq:Epol}
\Xi_I^\mathrm{pol} = 
\langle \Psi_{I:Q_I} | \hat{H}_{I:Q_I} | \Psi_{I:Q_I} \rangle -
\langle \Psi_I | \hat{H}_{I:Q_I} | \Psi_I \rangle,
\end{eqnarray}
the difference in the expectation of $\hat{H}_{I:Q_I}$ between the gas phase and in the embedding field,
and the Coulomb interaction energy between a molecule and the embedding field,
\begin{equation}
\label{eq:ECoul} 
    E_{I:Q_I}^\mathrm{Coul} = \langle \Psi_{I} | \hat{H}_{[I/Q_I]} | \Psi_{I}\rangle.
\end{equation}
such that $\Xi_I^\mathrm{elec} = \Xi_I^\mathrm{pol} + E_{I:Q_I}^\mathrm{Coul}$.
\citet{Hensen2004} further decomposed the polarization energy $\Xi_I^\mathrm{pol}$ into an energy of distorting the gas-phase wave function,
\begin{eqnarray} 
\label{eq:Edist}
    \Xi_I^\mathrm{dist} = \langle \Psi_{I:Q_I} | \hat{H}_{I} | \Psi_{I:Q_I} \rangle -
        \langle \Psi_I | \hat{H}_{I} | \Psi_I \rangle,
\end{eqnarray}
and the energy of stabilizing interactions with the embedding field $Q_I = \{ q_F \}$, 
\begin{eqnarray}
\label{eq:Estab}
    \Xi_I^\mathrm{stab}
        & = & \langle \Psi_{I:Q_I}| \hat{H}_{[I/Q_I]}|\Psi_{I:Q_I} \rangle
        - \langle \Psi_I | \hat{H}_{[I/Q_I]} | \Psi_I \rangle.
\end{eqnarray}

In a system of $N$ molecules, the total electronic interaction energy and its decomposition into polarization and permanent Coulomb energies are,
\begin{eqnarray}
\Xi^\mathrm{elec} &=& \Xi^\mathrm{pol} + E^\mathrm{Coul} \\
\Xi^\mathrm{pol} &=& \frac{1}{2} \sum_I \Xi_I^\mathrm{pol} \\
E^\mathrm{Coul} &=& \frac{1}{2} \sum_I E_{I:Q_I}^\mathrm{Coul}.
\end{eqnarray}
In the Coulomb and polarization energies, $E^\mathrm{Coul}$, the factor of 1/2 is introduced to compensate for doubly counting the interaction energy.
Like $\Xi^\mathrm{pol}$, $\Xi^\mathrm{dist}$ and $\Xi^\mathrm{stab}$ are similarly defined as sums over all molecules.
In our present scheme, only one molecule, the ligand, is treated quantum mechanically. Thus $\Xi^\mathrm{elec} = \Xi_I^\mathrm{elec}$, $\Xi^\mathrm{pol} = \Xi_I^\mathrm{pol}$, $\Xi^\mathrm{dist} = \Xi_I^\mathrm{dist}$, $\Xi^\mathrm{stab} = \Xi_I^\mathrm{stab}$, and $E^\mathrm{Coul} = E_{I:Q_I}^\mathrm{Coul}$, where $I$ is the ligand molecule.

 Both $\Psi_I$ and $\Psi_{I:Q_I}$ were calculated using the restricted Hartree-Fock method \cite{ostlund} in conjunction with the 6-311G** basis set \cite{6311gss}.
The atomic charge of atoms $A$ with and without the embedding field $Q_I = \{q_F\}$, 
$q_A^{\mathrm{QM}:Q_I}$ and $q_A^\mathrm{QM}$, respectively, 
were obtained by fitting to the quantum mechanical electrostatic potential (ESP) using the restrained electrostatic potential (RESP) method \cite{Bayly1993}. 

Fitted point charges were used to evaluate the stabilization energy,
\begin{eqnarray}
    \Xi^\mathrm{stab}
        & = & \sum_{A \in I} \sum_{F \in Q_I} \left(q_A^{\mathrm{QM}:Q_I}-q_A^{\mathrm{QM}} \right) 
        \frac{q_F}{R_{AF}}. \label{eq:stab_RESP}
\end{eqnarray}

For most reported calculations, the embedding field $Q_I = \{ q_F \}$ consisted of all of the non-ligand atoms in the system. In order to evaluate the distance at which embedding field atoms affect the polarization energy, we also performed calculations in which the embedding field consists of all atoms within a cutoff parameter $R_\mathrm{cut}$ of any ligand atom. The cutoff parameter was varied from $R_\mathrm{cut} \in \{ 4, 5, ..., 10, 12, .., 20 \}$. Even when different $R_\mathrm{cut}$ were used for determining $\Psi_{I:Q_I}$ and $q_A^{\mathrm{QM}:Q_I}$, energies were evaluated using an embedding field based on all atoms in the model.

In addition to the electrostatic interaction energy, coupling between the QM and MM region also includes a van der Waals interaction energy modeled by the Lennard-Jones potential,
\begin{eqnarray}
E^\mathrm{vdW} &=& \sum_{A \in I} \sum_{F \in Q_I}  
4\epsilon_{AF} 
\left[ \left( \frac{\sigma_{AF}}{R_{AF}}\right)^{12} - \left(\frac{\sigma_{AF}}{R_{AF}}\right)^{6} \right],
\end{eqnarray}
where $\sigma_{AF}$ and $\epsilon_{AF}$ are the Lennard-Jones parameters. Combined with $E^\mathrm{Coul}$, $E^\mathrm{vdW}$ makes up the intermolecular pairwise interaction energy,
\begin{eqnarray}
E^\mathrm{pair} = E^\mathrm{Coul} + E^\mathrm{vdW}.
\end{eqnarray}

For a system in solution, opposed to the gas phase, we also consider the solvation free energy. We will use $W(X)$ to denote a solvation free energy, where $X \in \{ PL, P, L \}$ represent the complex, protein, and ligand, respectively. The solvation free energy is an integral over all the solvent degrees of freedom. In principle, there are many ways to compute this quantity. In this paper, we used the Onufriev Bashford Case 2 (OBC2) \cite{Onufriev2004} generalized Born/surface area implicit solvent model.

\begin{figure}
	\includegraphics[width=3.33in]{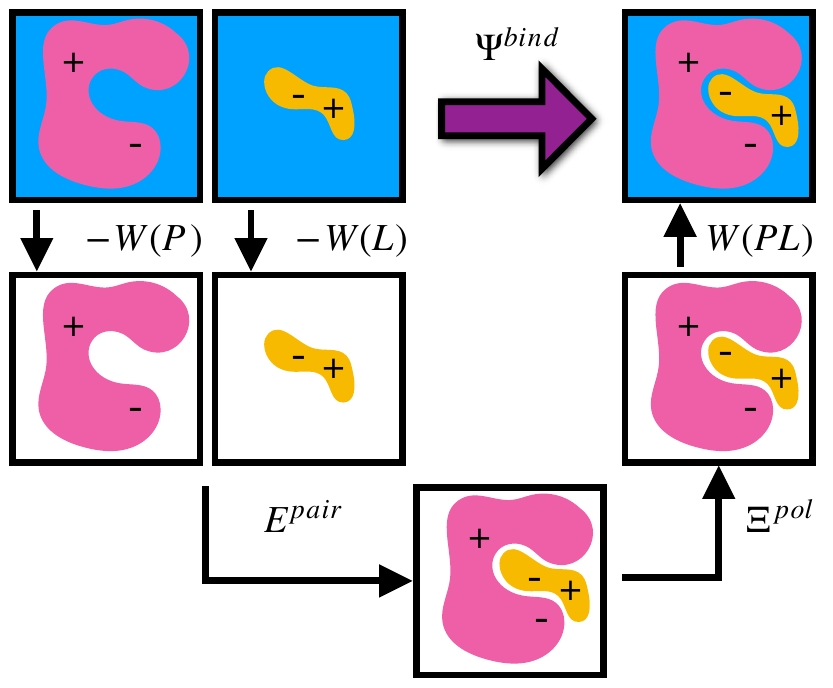}
	\caption{Schematic illustrating the decomposition of binding energy, $\Psi^\mathrm{bind}$,} into desolvation free energy of the protein, $-W(P)$, the desolvation free energy of the ligand, $-W(L)$, the intermolecular pairwise interaction energy, $E^\mathrm{pair}$, the ligand polarization energy, $\Xi^\mathrm{pol}$, and the solvation free energy of the complex, $W(PL)$.
	\label{fig:decomposition}
\end{figure}

The total binding energy is given by (Figure \ref{fig:decomposition}),
\begin{eqnarray}
\Psi^\mathrm{bind} &=& E^\mathrm{pair} + \Xi^\mathrm{pol} + W^\mathrm{bind}, \\
W^\mathrm{bind} &=& W(PL) - W(P) - W(L).
\end{eqnarray}
In the $W(PL)$ calculation, $q_A^{\mathrm{QM}:Q_I}$ are used for ligand partial charges. On the other hand, the $W(L)$ calculation uses $q_A^{\mathrm{QM}}$ for ligand partial charges.

To isolate the effects of polarization, we also define a total binding energy that does not consider ligand polarization,
\begin{eqnarray}
\Psi^\mathrm{bind, np} &=& E^\mathrm{pair} + W^\mathrm{bind, np}, \\
W^\mathrm{bind, np} &=& W(PL, np) - W(P) - W(L).
\end{eqnarray}
$W^\mathrm{bind, np}$ differs from $W^\mathrm{bind}$ because $q_A^{\mathrm{QM}}$ are used for ligand partial charges in the $W(PL)$ calculation. $\Psi^\mathrm{bind, np}$ is the binding energy for a purely MM model.


\subsection{Other Properties}

We computed a number of other properties to assess whether they have a clear relationship with the polarization energy. 

Motivated by the observation of a high density of acid and base side chains in enzymes \cite{Jimenez-Morales2012}, we computed two quantities: the percentage of atoms in a protein that are highly charged; and the number density of highly charged atoms within 6 \AA~ of any ligand atom. The percentage of atoms in the protein that are highly charged is defined as,
\begin{eqnarray}
\frac{1}{N} \sum_i^N H(|q_i|-0.6) \times 100,
\label{eq:perc_charged}
\end{eqnarray}
where $i$ is an index over atoms in the protein and $N$ is the total number of atoms in the protein. This expression uses the Heaviside step function,
\begin{eqnarray}
H(x) = 
\begin{cases}
0, & \quad  x < 0, \\
1, & \quad  x \geq 0,
\end{cases}
\end{eqnarray}
where $x$ is a real number. The volume of the binding site was determined by Monte Carlo integration. To perform this integration, a box was defined that includes 6~\AA~around the range of the ligand atoms in each dimension. Points within the box were randomly sampled from a uniform distribution and assessed for the distance to the nearest ligand atom. The site volume was estimated by the product of the box volume and the fraction of points in the box within 6~\AA~of a ligand atom.

We also computed a number of properties inspired by classical electrostatics. In classical electrostatics, the internal energy of a dipole moment in an electric field is the dot product of the dipole with the field. We considered two classical models: one in which the entire ligand is treated as a dipole and a second in which each atom is treated as a dipole. 

If the ligand is considered as a dipole, the change in internal energy due to an induced dipole is,
\begin{eqnarray}
\Xi^\mathrm{pol, cL} &=& 
- \pmb{\mu}_L^\mathrm{ind} \cdot \mathbf{E}_L^0, 
\label{eq:classical_polarization}
\end{eqnarray}
where $\pmb{\mu}_L^\mathrm{ind}$ is the induced dipole moment of the ligand $L$ 
and $\mathbf{E}_L^0$ is the electric field acting on the ligand $L$ due to the embedding field $Q_L = \{ q_F\}$ consisting of atomic charges of the surrounding atoms. The electric field acting on the center of mass (or protons) $\pmb{R}_C$ of the ligand is,
\begin{eqnarray}
\mathbf{E}_{L}^0 = \sum_{F \in Q_L}  \frac{q_F}{R_{CF}^3} \pmb{R}_{CF},
\label{eq:efield_center}
\end{eqnarray}
where  $F$ runs over the atomic sites in the embedding field. $\pmb{R}_{CF} = \pmb{R}_C - \pmb{R}_F$ and $R_{CF} = | \pmb{R}_{CF} |$.

The induced dipole moment of the ligand $\pmb{\mu}_L^\mathrm{ind}$ was calculated in two ways. The first was from the expectation value of the dipole moments,
\begin{eqnarray}
\pmb{\mu}_L^\mathrm{ind,\mathrm{QM}} &=&
\langle \Psi_{L:Q_L}|\hat{\mu}|\Psi_{L:Q_L}\rangle
- \langle \Psi_{L}|\hat{\mu}|\Psi_{L}\rangle \nonumber \\
& = & \pmb{\mu}_L^{\mathrm{QM}:Q_L} - \pmb{\mu}_L^\mathrm{QM},
\label{eq:induced_dipole_QM}
\end{eqnarray}
where $\hat{\mu}$ is the dipole moment operator. The second was based on the molecular polarizability tensor, $\pmb{\alpha}_L$, and the electric field on the center of mass of the ligand,
\begin{eqnarray}
\pmb{\mu}_{L}^\mathrm{ind, \alpha_L} & = & \pmb{\alpha}_L \mathbf{E}_{L}^0.
\label{eq:induced_dipole_tensor_com}
\end{eqnarray}
Elements of the molecular polarizability tensor $(\pmb{\alpha}_L)_{xy}$ describe the susceptibility of a molecule to polarization along the $x$ axis due to an electric field along the $y$ axis. As in \citet{Willow2015}, these tensor elements were calculated based on placing a pair of point charges of $\mp$1 a.u. at $\bm{R}_\mathrm{cm} \pm 100$ Bohr along a Cartesian axis, where $\bm{R}_\mathrm{cm}$ represents the center of mass of the ligand, to create an electric field. Then $(\pmb{\alpha}_L)_{xy}$ were evaluated as the ratio of the induced dipole moment due the point charges, $\pmb{\mu}_L^\mathrm{ind, pc}$, and the electric field applied by the point charge onto the ligand, $\mathbf{E}_L^\mathrm{0, pc}$, 
\begin{eqnarray}
(\pmb{\alpha}_L)_{xy} &=& \frac{ ( \pmb{\mu}_L^\mathrm{ind, pc})_x } {(\mathbf{E}_L^\mathrm{0, pc})_y}.
\end{eqnarray}
The dipole moment from the electron density is more accurate and valuable for assessing the correspondence between $\Xi^\mathrm{pol}$ and $\Xi^\mathrm{pol, c}$. However, it is not a practical shortcut to the polarization energy because it requires the same quantum chemistry calculations used to compute $\Xi^\mathrm{pol}$. On the other hand, although the molecular polarizability tensor, $\pmb{\alpha}_L$, requires three quantum chemistry calculations, it can be reused (as an approximation) for multiple ligand configurations. Hence, the dipole moment from the molecular polarizability tensor, $\pmb{\mu}_L^\mathrm{ind, \alpha_L}$, could potentially reduce the computational costs of $\Xi^\mathrm{pol}$ prediction. To facilitate comparison with the polarization energy, we also computed the molecular polarizability scalar of the ligand, $\alpha_L$, defined as,
\begin{eqnarray}
\alpha_L = \frac{1}{3} \mathrm{Tr}[ \pmb{\alpha}_L ],
\label{eq:pol_scalar}
\end{eqnarray}
where $\mathrm{Tr}$ is the trace of a square matrix.

If each atom on the ligand is considered as a dipole, then the change in internal energy due to an induced dipole is,
\begin{eqnarray}
\Xi^\mathrm{pol, cA} &=& 
- \sum_{A \in L} \pmb{\mu}_A^\mathrm{ind} \cdot \mathbf{E}_A^0, 
\label{eq:classical_polarization_atom}
\end{eqnarray}
The electric field acting on an atom is,
\begin{eqnarray}
\mathbf{E}_{A}^0 = \sum_{F \in Q_L} \frac{q_F}{R_{AF}^3} \pmb{R}_{AF},
\label{eq:efield_atom}
\end{eqnarray}
where $A$ runs over all atomic sites in the ligand. 
The induced dipole on each atom was computed based on RESP charges as,
\begin{eqnarray}
\pmb{\mu}_A^\mathrm{ind} = (q_A^\mathrm{QM:Q_L} - q_A^\mathrm{QM}) \pmb{R}_A.
\end{eqnarray}

In all, we considered the relationship between $\Xi^\mathrm{pol}$ and a number of other properties: the
\begin{enumerate}
    \item percentage of highly charged atoms in a protein (Eq. \ref{eq:perc_charged}); 
    \item molecular polarizability scalar, $\alpha_L$ (Eq. \ref{eq:pol_scalar});
    \item Coulomb interaction energy, $E^\mathrm{Coul}$ (Eq. \ref{eq:ECoul});
    \item magnitude of the electric field on the ligand center of mass, $|\mathbf{E}_L^0|$, where $\mathbf{E}_L^0$ is from Eq. \ref{eq:efield_center};
    \item magnitude of total electric field on the ligand atom sites, $\left|\sum_{A \in L} \mathbf{E}_{A}^0 \right|$, where $\mathbf{E}_{A}^0$ is from Eq. \ref{eq:efield_atom};
    \item magnitude of the induced dipole moment based on wave functions, $|\pmb{\mu}_L^\mathrm{ind, QM}|$, where $\pmb{\mu}_L^\mathrm{ind,\mathrm{QM}}$ is from Eq. \ref{eq:induced_dipole_QM};
    \item magnitude of the induced dipole moment based on the molecular polarizability tensor, $|\pmb{\mu}_L^\mathrm{ind, \alpha_L}|$, where $\pmb{\mu}_L^{ind, \alpha_L}$ is from Eq. \ref{eq:induced_dipole_tensor_com};
    \item classical polarization energy of a ligand dipole, $\Xi^\mathrm{pol, cL}$ (Eq. \ref{eq:classical_polarization}), using Eq. \ref{eq:induced_dipole_QM} for the induced dipole moment.
    \item classical polarization energy of a ligand dipole, $\Xi^\mathrm{pol, cL, \alpha_L}$ (Eq. \ref{eq:classical_polarization}), using Eq. \ref{eq:induced_dipole_tensor_com} for the induced dipole moment.
    \item and classical polarization energy of atomic dipoles, $\Xi^\mathrm{pol, cA}$ (Eq. \ref{eq:classical_polarization_atom}).
\end{enumerate}

\subsection{Computational Methods}

Structures from the PDBBind Core Set (release 2016) were processed through an automated workflow based on AmberTools 17 \cite{Case2017} and customized QM/MM codes. Protein protonation states were assigned using PDB2PQR 1.9.0 at a pH of 7.0 and ligand protonation states using pkatyper in the QUACPAC 1.7.0.2 toolkit (OpenEye). AMBER topology files based on protein and cation parameters (Na$^+$, Mg$^{2+}$, Ca$^{2+}$, and Zn$^{2+}$) from the AMBER ff14SB force field \cite{Maier2015} and ligand parameters from the Generalized AMBER Force Field 2 \cite{Wang2006} were built using AmberTools 17 \cite{Case2017}.

Using OpenMM 7.3.1\cite{Eastman2010}, complexes in OBC2 \cite{Onufriev2004} generalized Born/surface area implicit solvent were minimized with heavy atom restraints of 2 kcal/mol/\AA$^2$ towards crystallographic positions until energies converged within 0.24 kcal/mol.

In our modified QM/MM codes, the evaluation of molecular integrals of many-body operators over Gaussian functions were obtained using libint 2.5.0 \cite{Libint2} and the linear algebra and eigenvalue decomposition of a symmetric matrix were done with the Armadillo 8.500.1 \cite{Sanderson2016,Sanderson2018}.

OpenMM 7.3.1\cite{Eastman2010} was also used to evaluate van der Waals and solvation energies, the latter with the OBC2 \cite{Onufriev2004} generalized Born/surface area implicit solvent model.

\section{Results and Discussion}

\subsection{The distribution of polarization energy is broad and skewed}

\begin{figure}
	\includegraphics[width=0.45\linewidth]{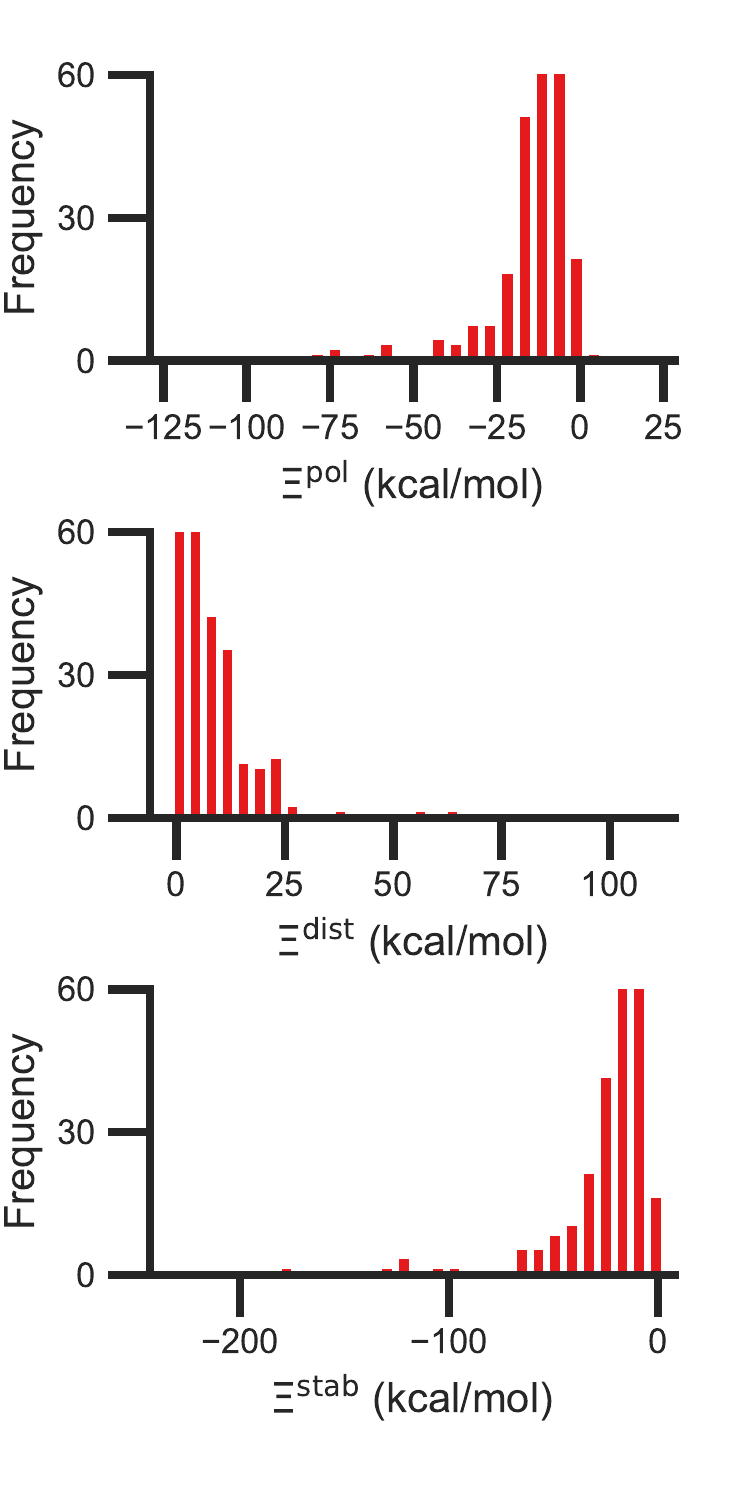}
	\caption{Histograms of the ligand polarization (top, $\Xi^\mathrm{pol}$), distortion (middle, $\Xi^\mathrm{dist}$), and stabilization (bottom, $\Xi^\mathrm{stab}$) energies in the PDBBind Core Set. The three quantities are related by $\Xi^\mathrm{pol} = \Xi^\mathrm{dist} + \Xi^\mathrm{stab}$.}
	\label{fig:hist_Xipol}
\end{figure}

Signs of the calculated polarization energy $\Xi^\mathrm{pol}$, the distortion energy $\Xi^\mathrm{dist}$, and the stabilization energy $\Xi^\mathrm{stab}$ are mostly as expected (Fig. \ref{fig:hist_Xipol}). In nearly all of the calculations, $\Xi^\mathrm{pol} < 0$, $\Xi^\mathrm{dist} > 0$, and $\Xi^\mathrm{stab} < 0$. The embedding field reshapes the wave function to have stronger Coulomb interactions between the electronic probability density and point charges, such that $\Xi^\mathrm{stab} < 0$. Because the gas-phase wave function of the ligand has the optimal intramolecular potential, perturbing the wave function leads to a higher intramolecular potential energy such that $\Xi^\mathrm{dist} > 0$. In the vast majority of systems, the calculated distortion is more than compensated for by the calculated stabilization such that the calculated net effect on the interaction energy due to polarization, $\Xi^\mathrm{pol}$, is negative.

Exceptions to the trend of negative calculated ligand polarization energies are due to structural modeling issues that lead to short intermolecular distances. Positive $\Xi^\mathrm{pol}$ values were calculated in three complexes. In our models of these structures, there are very short distances between a hydrogen atom in the ligand and in the protein: 0.73 \AA~in 2fxs, 1.06 \AA~in 3u5j, and 0.87 \AA~in 4f2w. The close proximity of atoms leads to a severe distortion in the wave function that is not overcome by more favorable Coulomb interactions. These steric clashes could be resolved by changing the models in minor ways that are equally compatible with crystallographic evidence and pKa predictions. In the 2fxs and 4f2w models, the proton on a carboxylic acid was arbitrarily placed near a ligand hydrogen instead of on the other carbonyl oxygen. In the 3u5j model, the clash could be resolved by switching the position of the terminal oxygen and amine groups, which have nearly identical electron density, on asparagine 140.

The distribution of $\Xi^\mathrm{pol}$, $\Xi^\mathrm{dist}$, and $\Xi^\mathrm{stab}$ is broad and skewed. There is a peak in the distribution of $\Xi^\mathrm{pol}$ around -10 kcal/mol. However, for a small number of complexes, $\Xi^\mathrm{pol}$ is much lower, with a minimum value of -128 kcal/mol.

\subsection{Systems with the lowest $\Xi^\mathrm{pol}$ have close cations}

We hypothesized that the lowest $\Xi^\mathrm{pol}$ could be due to crystallographic cations. To test this hypothesis, we subdivided the PDBBind Core Set into two subsets: 90 complexes with cations (Na$^+$, Mg$^{2+}$, Ca$^{2+}$, and Zn$^{2+}$) and 196 complexes without cations in the crystal structure. 

Histograms of $\Xi^\mathrm{pol}$ for the two subsets are consistent with our hypothesis (Fig. S1 in the Supporting Information). All systems in which $\Xi^\mathrm{pol}$ < -50 kcal/mol are in the subset with cations. In contrast, the minimum $\Xi^\mathrm{pol}$ in the subset without cations is around -40 kcal/mol. The range of $\Xi^\mathrm{dist}$ and $\Xi^\mathrm{stab}$ is also much smaller in the subset without cations.

Crystallographic cations may have an outsize role in ligand polarization because the magnitude of their charge is larger than the charge of most protein atoms. In the AMBER ff14SB force field \cite{Maier2015}, protein partial charges were determined by applying RESP \cite{Bayly1993} to electrostatic potentials from QM calculations. Most protein atoms have near-zero charge. The magnitude of the charge is greater than 0.6$e$, where $e$ represents the elementary charge, in only a few atoms. It is less than 1$e$ in all atoms. These conclusions are also true for protein atoms in our data set (Fig. S2 in the Supporting Information). The low magnitude of charge results from delocalization of net charges across several atoms. In contrast, the cations have a charge of +1$e$ or +2$e$ that is localized onto a single atom and have a more focused effect on the electrostatic potential.

\begin{figure}
	\includegraphics[width=0.45\linewidth]{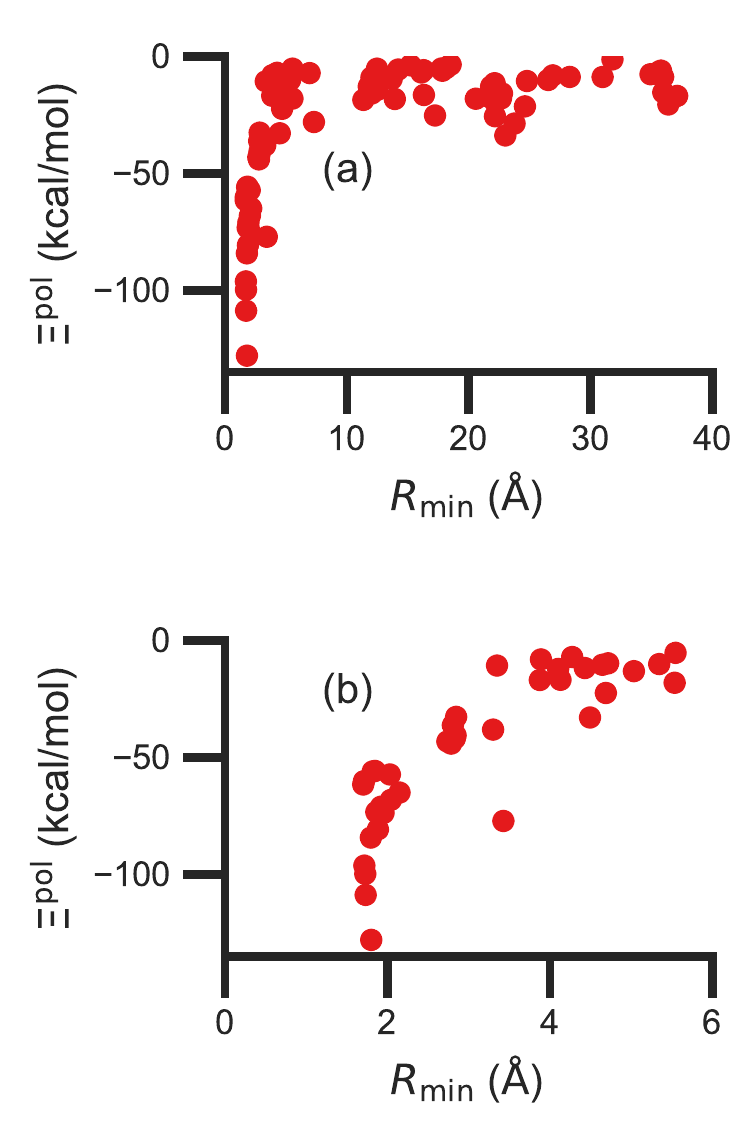}
	\caption{Scatter plot of the ligand polarization energy $\Xi^\mathrm{pol}$ as a function of the minimum distance between a ligand and cation atom, $R_\mathrm{min}$, for (a) the entire range of $R_\mathrm{min}$ and (b) $R_\mathrm{min}$ < 6 \AA.
	\label{fig:scatter_Xipol_Rmin}}
\end{figure}

Beyond the presence of cations, the distance between ligand and cation atoms also plays an important role in ligand polarization (Fig. \ref{fig:scatter_Xipol_Rmin}). Even if cations are present in a crystal structure, they are not necessarily close enough to the ligand to significantly polarize its wave function. In many systems, cations are over 10 \AA~from any ligand atom. In all of the complexes in which $\Xi^\mathrm{pol}$ < -50 kcal/mol, a cation is within 4 \AA~of a ligand atom.

\begin{figure}
	\begin{tabular}{c c}
		\includegraphics[width=0.4\linewidth]{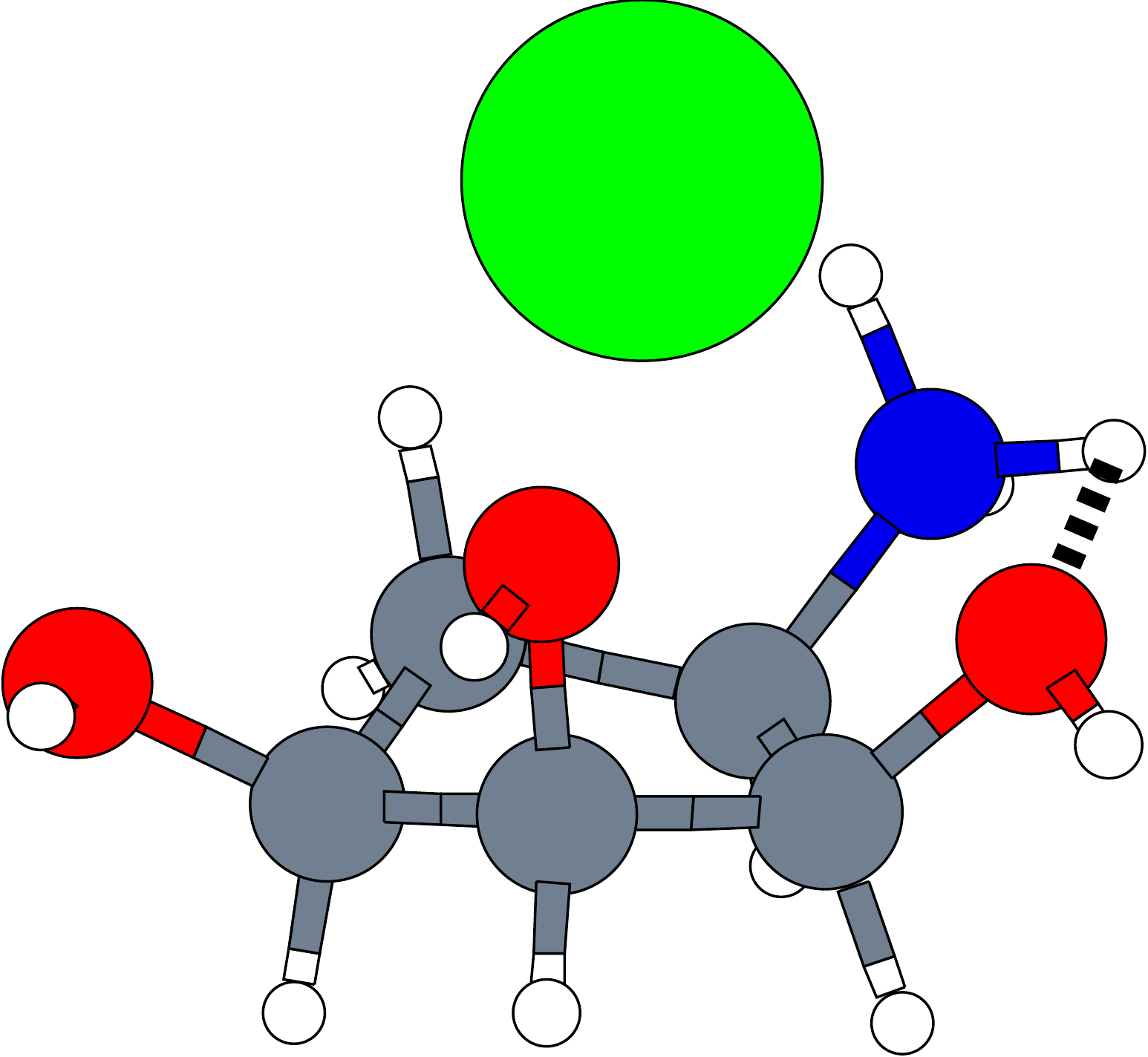} &
		\includegraphics[width=0.4\linewidth]{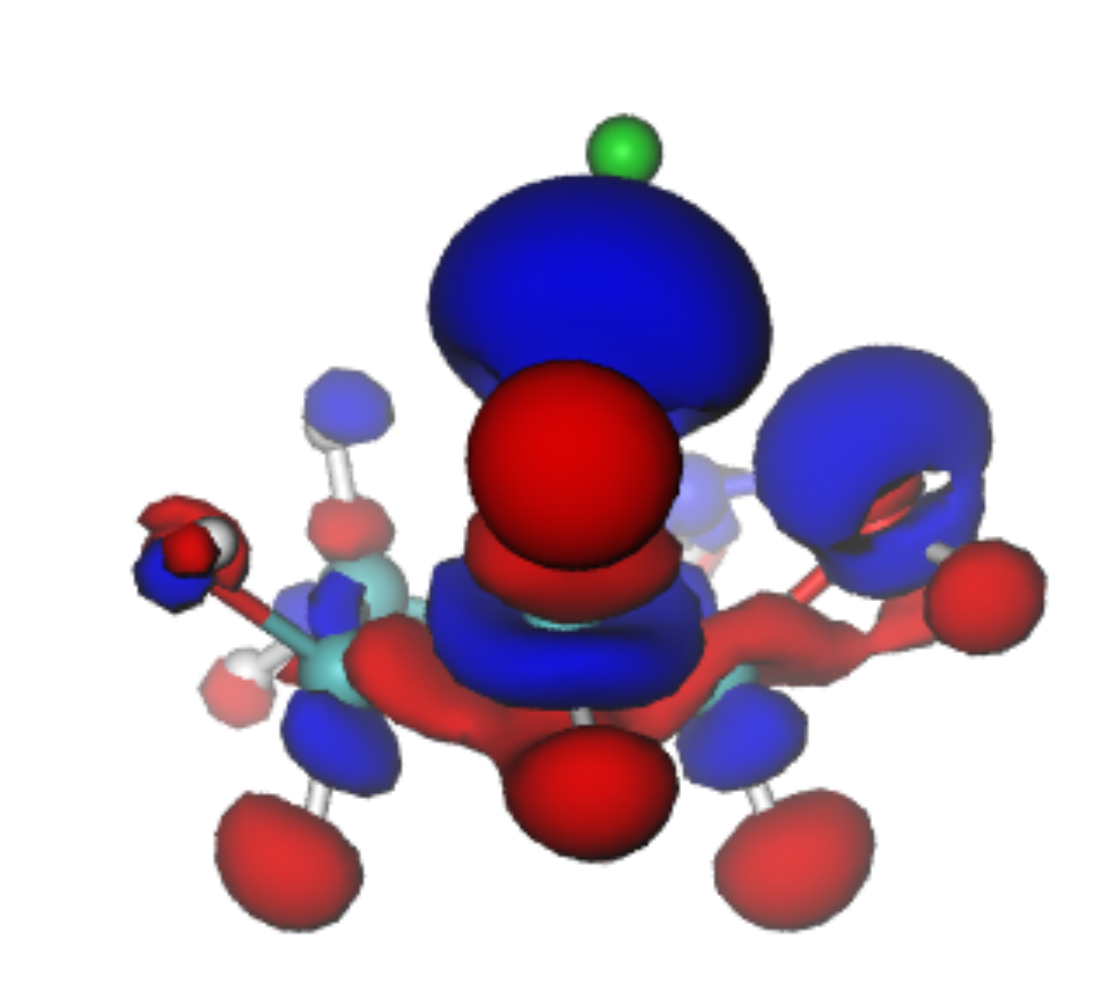} \\
		(a) & (b) 
	\end{tabular}
	\caption{(a) The molecular structure of the ligand with one Zinc cation Zn$^{2+}$ in the complex 3dx1. Hydrogen, carbon, nitrogen, oxygen, and zinc atoms are colored with white, gray, blue, red, and green, respectively. 
	(b) The difference in the electronic probability density is plotted. Blue and red contours illustrate the gain and loss of the electronic probability density due to the embedding field.}
	\label{fig:3dx1}
\end{figure}

Unfortunately, the extent of ligand polarization when ligands are close to cations is likely overestimated by our QM/MM scheme. Because only the ligand is included in the QM region, cations are simply represented as positive point charges. While actual cations have inner-shell electrons that repel further electron density, the point charges are purely attractive. The purely attractive forces draw an unrealistic amount of electron density between the ligand and cation, leading to a very negative polarization energy.  For an estimate of the extent of overpolarization in several systems, see Table S1 in the Supporting Information. As an illustrative example, there is a significant gain in the electron density between the ligand and cation in the complex 3dx1 (Fig. \ref{fig:3dx1}). Hence, we will proceed with extra caution in interpreting points where $\Xi^\mathrm{pol}$ < -50 kcal/mol.

\subsection{The importance of  the embedding field size diminishes with distance}

\begin{figure}
	\includegraphics[width=0.9\linewidth]{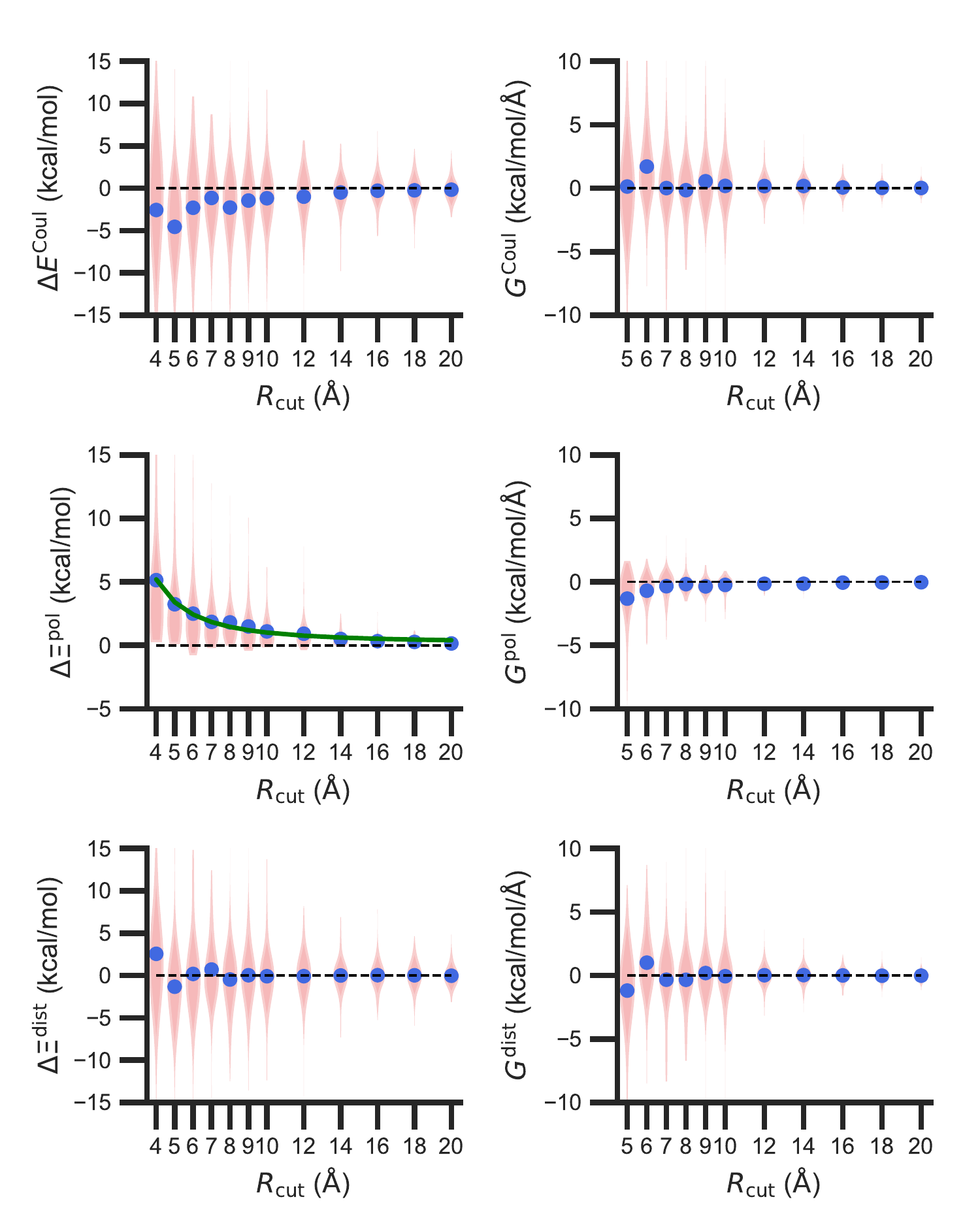}
	\caption{Dependence of the Coulomb interaction $E^\mathrm{Coul}$, the ligand polarization energy $\Xi^\mathrm{pol}$, and the distortion energy $\Xi^\mathrm{dist}$ on the cutoff distance $R_\mathrm{cut}$. Here, the deviation and the gradient are defined as $\Delta F(R_\mathrm{cut}) = F(R_\mathrm{cut}) - F(\infty)$ and $G = d F(R_\mathrm{cut}) / d R_\mathrm{cut}$, respectively, where $F$ is either $E$ or $\Xi$. In these violin plots, the width of the shaded area is proportional to the frequency of observations. Large blue points are placed at mean values. In the plot of $\Delta \Xi^\mathrm{pol}$ as a function of $R_\mathrm{cut}$, the green line is a function that was fitted to the mean values, $80.778 R_\mathrm{cut}^{-2} + 0.177$.}
	\label{fig:violin_DeltaXipol_Rcut}
\end{figure}

The size of the embedding field strongly affects estimates of the polarization energy (Fig. \ref{fig:violin_DeltaXipol_Rcut}). Changes in the cutoff distance $R_\mathrm{cut}$ alter the partial charges included in the embedding field, the wave function $\Psi_{I:Q_I}$, and then the RESP charges. Regardless of $R_\mathrm{cut}$, nearly every estimate of 
$\Delta \Xi^\mathrm{pol}(R_\mathrm{cut}) = \Xi^\mathrm{pol}(R_\mathrm{cut}) - \Xi^\mathrm{pol}(R_\mathrm{cut}=\infty)$ is positive, indicating that the ligand wave function accommodates even distant charges in the embedding field. However, the influence of protein and cation charges diminishes with distance. Correspondingly, as $\Delta \Xi^\mathrm{pol}$ diminishes, so does its variance. For larger values of $R_\mathrm{cut} =$ 8, 9, 10, and 12 \AA, the mean (and standard deviation) of $\Delta \Xi^\mathrm{pol}$ is  1.81 (1.77), 1.49 (1.80), 1.10 (1.23), and 0.92 (1.14) kcal/mol, respectively.

The decomposition of the polarization energy into $E^\mathrm{Coul}$ and $E^\mathrm{dist}$ is more sensitive to $R_\mathrm{cut}$ than the polarization energy itself; distributions of the values (relative to values with no cutoff) and numerical derivatives are broader. Even at $R_\mathrm{cut} =$ 8, 9, 10, and 12 \AA, the mean (and standard deviations) of $\Delta E^\mathrm{Coul}$ are -2.28 (4.94), -1.46 (4.71), -1.18 (4.06), and -0.99 (3.38) kcal/mol.

On average, the decay of $\Delta \Xi^\mathrm{pol}$ is well-described by an inverse square law. 
A nonlinear least-squares regression using scipy.optimize.curve\_fit (\url{https://scipy.org/}) of $x_1 R_\mathrm{cut}^{-2} + x_2$ for $x_1$ and $x_2$ yielded a curve that closely matches the data. The curve is best for low $R_\mathrm{cut}$, slightly underestimates the mean for intermediate $R_\mathrm{cut}$, and slightly overestimates the mean for larger $R_\mathrm{cut}$. The inverse square power law is consistent with the $R^{-4}$ dependence of ion-induced dipole interactions because the volume of the region containing embedding field charges increases as $R_\mathrm{cut}^2$.

\subsection{Of computed properties, $\Xi^\mathrm{pol}$ is most correlated with the electric field, the induced dipole moment, and the classical polarization energy}

We observed that a number of properties - the percentage of atoms in a protein that are highly charged, the number density of highly charged atoms, and the Coulomb interaction energy - have little or only weak correlation with the ligand polarization energy (Figure S3 in the Supporting Information).

\begin{figure} 
	\includegraphics[width=0.9\linewidth] {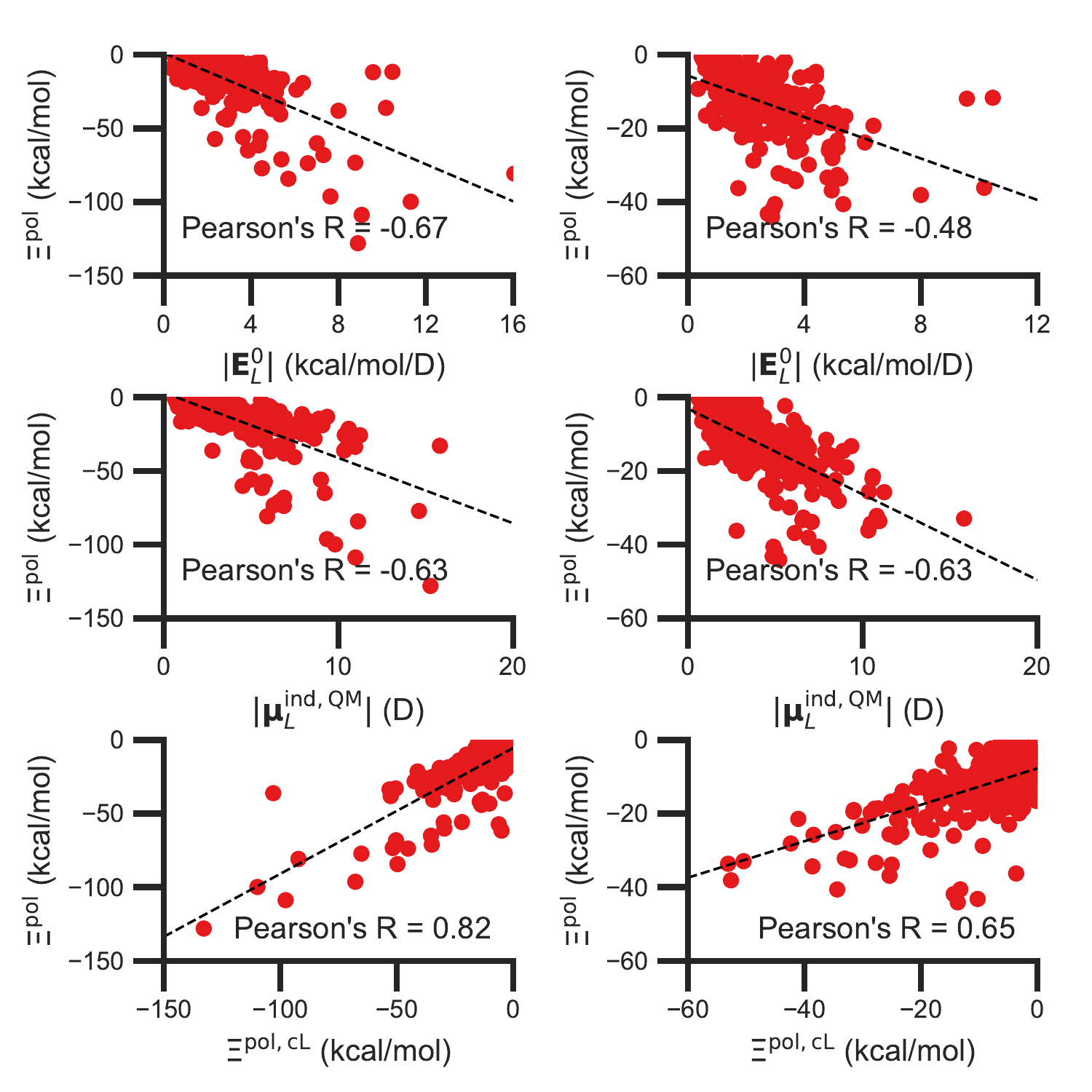}
\caption{The ligand polarization energy, $\Xi^\mathrm{pol}$, as a function of the magnitude of the electric field $|\mathbf{E}_L^0|$ (top), 
the magnitude of the induced dipole moment $|\pmb{\mu}_L^\mathrm{ind, QM}|$ (middle), 
and the classical polarization energy $\Xi^\mathrm{pol, cL}$ (bottom),
where $\mathbf{E}_L^0$, $\pmb{\mu}_L^\mathrm{ind,\mathrm{QM}}$, and $\Xi^\mathrm{pol, cL}$ are from Eq. \ref{eq:efield_center}, Eq. \ref{eq:induced_dipole_QM}, and Eq. \ref{eq:classical_polarization}, respectively.
The range of $\Xi^\mathrm{pol}$ is either $\Xi^\mathrm{pol} < 0$ kcal/mol (left) or -50 kcal/mol $< \Xi^\mathrm{pol} < 0$ kcal/mol (right).
\label{fig:scatter_Xipol_summary}}
\end{figure}

In contrast with the aforementioned properties, there is a much clearer relationship between the ligand polarization energy, $\Xi^\mathrm{pol}$, and several other properties:
the magnitude of the electric field; 
the magnitude of the induced dipole moment of the ligand;
and the classical polarization energy (Fig. \ref{fig:scatter_Xipol_summary}).
The linear correlation is strong with the magnitude of the electric field on the ligand center of mass, $|\mathbf{E}_{L}^0|$, and even stronger with the magnitude of the total electric field vector active on all ligand atoms, $\left|\sum_{A \in L} \mathbf{E}_{A}^0 \right|$ (Fig. \ref{fig:scatter_Xipol_summary} and S5 in the Supporting Information).
Intriguingly, in both cases, there appear to be two distinct trends relating the electric field to the magnitude of the electric field; a linear correlation exists in systems where $\Xi^\mathrm{pol} < $ -50 kcal/mol, but the slope is distinct from in systems where  -50 kcal/mol $< \Xi^\mathrm{pol} < $ 0 kcal/mol.
The two measures of the electric field are also correlated with each other, with a Pearson's R of 0.54 (Fig. S6 in the Supporting Information).
Similarly, the ligand polarization energy $\Xi^\mathrm{pol}$ is also strongly correlated with the magnitude of the induced dipole moment of the ligand. There is a stronger correlation with the magnitude of the induced dipole moment based on wave functions $|\pmb{\mu}_L^\mathrm{ind, QM}|$, where $\pmb{\mu}_L^\mathrm{ind,\mathrm{QM}}$ is from Eq. \ref{eq:induced_dipole_QM}, than the magnitude of the induced dipole moment based on the molecular polarizability tensor, $|\pmb{\mu}_L^\mathrm{ind, \alpha_L}|$, where $\pmb{\mu}_L^{ind, \alpha_L}$ is from Eq. \ref{eq:induced_dipole_tensor_com} (Fig. \ref{fig:scatter_Xipol_summary} and S7 in the Supporting Information).

Finally, in addition to the strong relationship between the ligand polarization energy $\Xi^\mathrm{pol}$ and both the magnitude of the electric field and the induced dipole, there is also a clear correspondence between the ligand polarization energy $\Xi^\mathrm{pol}$ and the classical polarization energy. Of approaches to compute the classical polarization energy, treating the entire ligand as a dipole and using Eq. \ref{eq:induced_dipole_QM} for the induced dipole moment led to the best correlation with the quantum polarization energy (Fig. \ref{fig:scatter_Xipol_summary} and S8 in the Supporting Information).
The clear correlation between the two quantities suggests that the classical model of a dipole in an electric field is a reasonable explanation for the quantum behavior. Limitations of the molecular polarizability model are described in Fig. S9 and S10 in the Supporting Information.

The observed linear correlation between the ligand polarization energy and the magnitude of the electric field $|\mathbf{E}_L^0|$ (Fig. \ref{fig:scatter_Xipol_summary}) has potential implications for modeling protein-ligand interactions with MM, including molecular docking. Because $|\mathbf{E}_L^0|$ is computed without a QM calculation, a relatively inexpensive polarization energy estimate based on linear regression can be added to binding energy estimates. Such an approach could recapitulate some of the success of semi-empirical QM in reconstructing binding poses \cite{Pecina2017,Ajani2017,Pecina2016,Chaskar2014}.

\subsection{Polarization is a substantial and variable fraction of interaction and binding energies} 

\begin{figure}
	\includegraphics[width=0.45\linewidth]{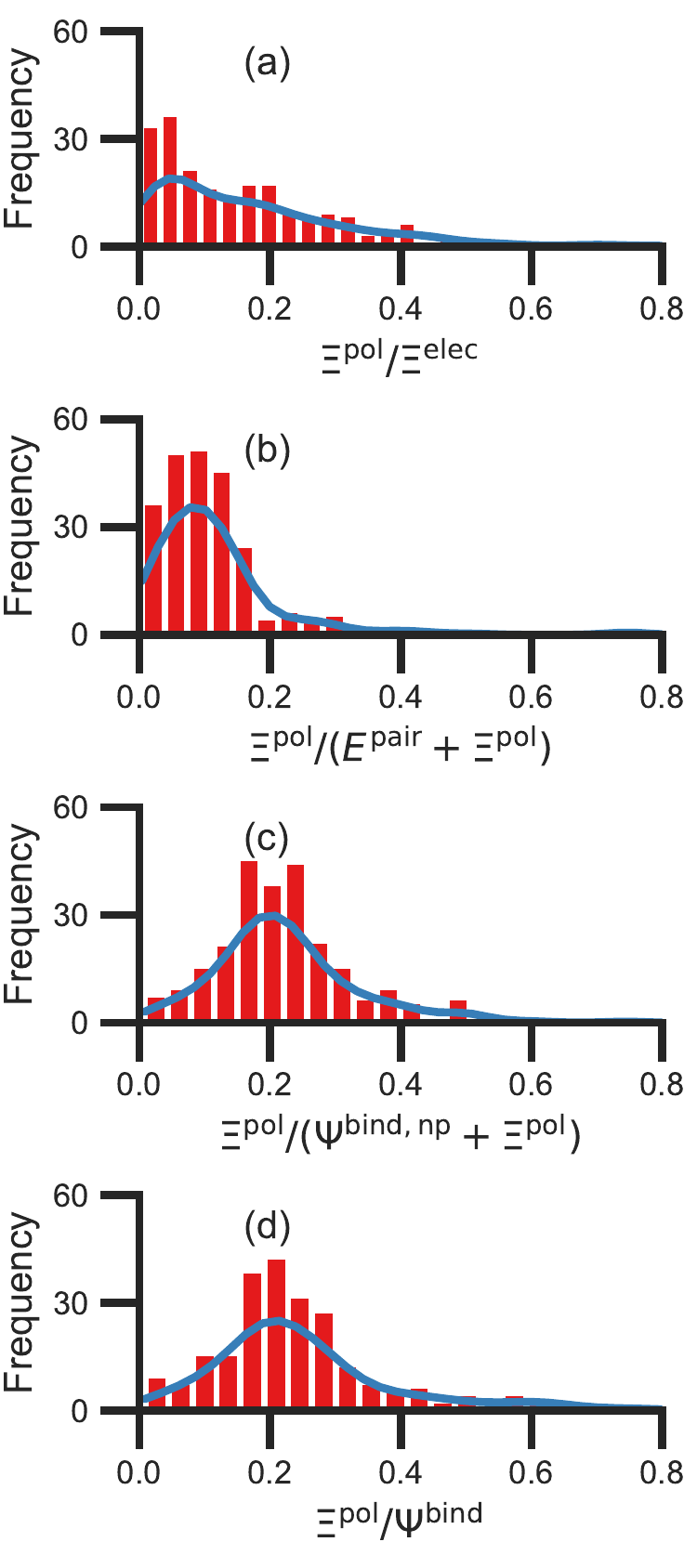}
	\caption{Histograms of ratio of the polarization energy of the ligand to (a) the electrostatic interaction ($\Xi^\mathrm{elec} = E^\mathrm{Coul} +\Xi^\mathrm{pol} $), (b) the intermolecular pairwise potential energy with the ligand polarization energy ($E^\mathrm{pair} + \Xi^\mathrm{pol}$), (c)  the  binding energy without considering ligand polarization in the solvation free energy ($\Psi^\mathrm{bind,np} + \Xi^\mathrm{pol}$), and (d) the binding energy with considering ligand polarization in the solvation free energy ($\Psi^\mathrm{bind}$).
	The histograms are truncated at a ratio of 1.25. Data are only included for complexes where $\Xi^\mathrm{pol} < 0$ kcal/mol (left) or -50 kcal/mol $< \Xi^\mathrm{pol} < 0$ kcal/mol. For analogous histograms including all data, see Fig. S11 in the Supporting Information.
	\label{fig:hist_ratio}
}
\end{figure}

We observe that the ligand polarization energy $\Xi^\mathrm{pol}$ can be a substantial and highly system-dependent fraction of the interaction energy and binding energy (Fig. \ref{fig:hist_ratio}). In most systems where -50 kcal/mol $ < \Xi^\mathrm{pol} < 0$ kcal/mol, the ratio $\Xi^\mathrm{pol}/\Xi^\mathrm{elec}$ ranges from 0 to 0.4 (Fig. \ref{fig:hist_ratio}a). Exceptions occur when $\Xi^\mathrm{elec}$ is positive, leading to a negative ratio, or when it is small, leading to a ratio much larger than 1 (Table S2 in the Supporting Information). Positive and small values of $\Xi^\mathrm{elec}$ result from positive $E^\mathrm{Coul}$. For example, the complex 5c2h has $\Xi^\mathrm{pol} = -22.45$ kcal/mol, $E^\mathrm{Coul} = 19.60$ kcal/mol, and $\Xi^\mathrm{elec} = -2.85$ kcal/mol. Hence, $\Xi^\mathrm{pol}/\Xi^\mathrm{elec}= -7.88$. The histogram of $\Xi^\mathrm{pol}/(E^\mathrm{pair} + \Xi^\mathrm{pol})$ is compressed compared to $\Xi^\mathrm{pol}/\Xi^\mathrm{elec}$, with the range with the largest density reduced to between 0 and 0.2 (Fig. \ref{fig:hist_ratio}b). Smaller ratios are due to the addition of van der Waals interactions that increase values in the denominator. The histograms of $\Xi^\mathrm{pol}/(\Psi^\mathrm{bind, np} + \Xi^\mathrm{pol})$ and $\Xi^\mathrm{pol}/\Psi^\mathrm{bind}$ is notable for a clear peak around 0.2 (Fig. \ref{fig:hist_ratio}c \& d). If all systems in the PDBBind are considered, qualitative trends are similar but there is increased density at higher ratios (Fig. S11 in the Supporting Information).

When considering the polarization energies of three HIV-protease inhibitors, \citet{Hensen2004} found that $\Xi^\mathrm{pol}$ can approach one-third of the electrostatic interaction energy. In our much larger data set, we found that $\Xi^\mathrm{pol}$ can be a larger fraction of $\Xi^\mathrm{elec}$.

\begin{figure}
	\includegraphics[width=0.45\linewidth] {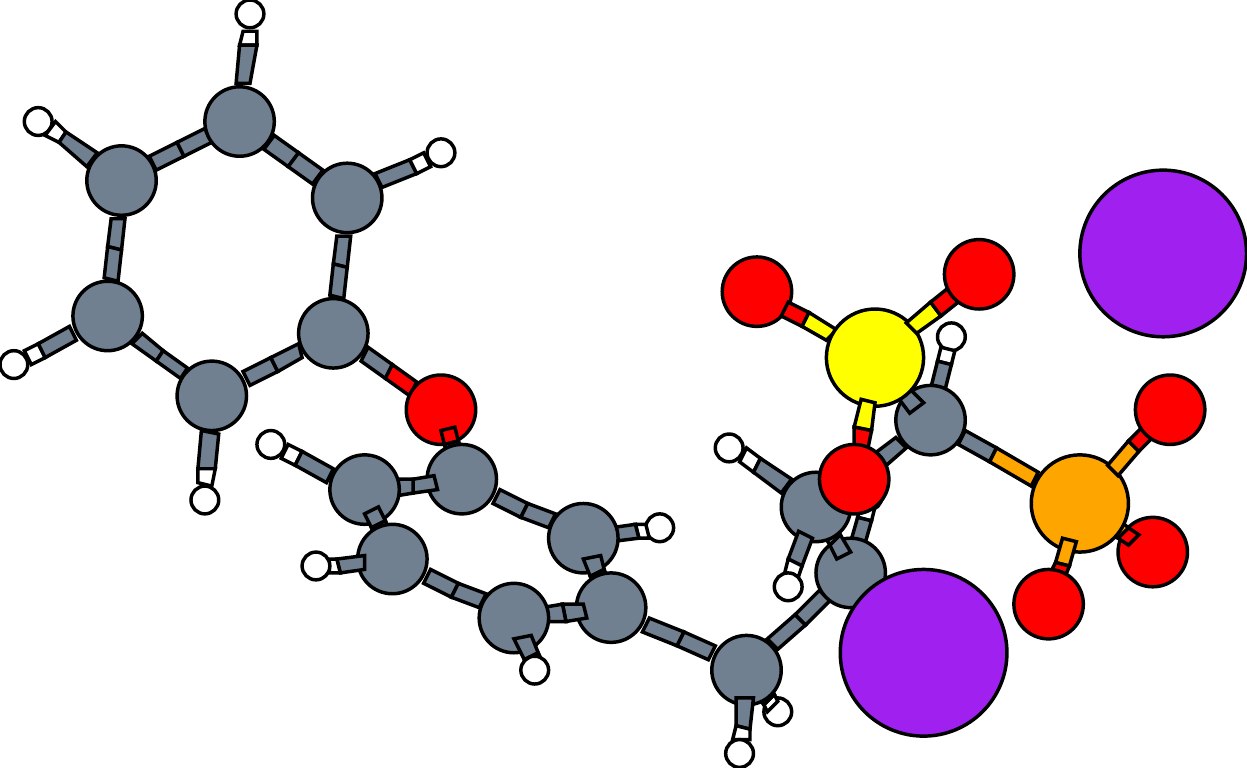}
	\caption{The molecular structure of the ligand with two Magnesium cations Mg$^{2+}$ in the complex, 2zcq. Hydrogen, carbon, oxygen, magnesium, phosphorus, and sulfur atoms are colored with white, gray, red, pink, orange, and yellow, respectively.}
	\label{fig:2zcq}
\end{figure}

With the caveat that polarization could be overestimated in these cases, two examples where $\Xi^\mathrm{pol}/(\Psi^\mathrm{bind,np}+\Xi^\mathrm{pol})$ is particularly large, 3dx1 and 3dx2, highlight the potentially outsized importance of $\Xi^\mathrm{pol}$ for small ligands (Table  S2 in the Supporting Information). 
In 3dx1, $\Psi^\mathrm{bind,np} + \Xi^\mathrm{pol} = -12.953$ kcal/mol, 
$\Xi^\mathrm{pol} = -80.77$ kcal/mol, 
and the ratio is $\Xi^\mathrm{pol}/(\Psi^\mathrm{bind,np}+\Xi^\mathrm{pol}) = 6.236$. 
For comparison, in 2zcq, $\Psi^\mathrm{bind,np}+\Xi^\mathrm{pol} = -295.48$ kcal/mol, $\Xi^\mathrm{pol} = -128.01$ kcal/mol, and the ratio is $\Xi^\mathrm{pol}/(\Psi^\mathrm{bind,np}+\Xi^\mathrm{pol}) = 0.43$. The ligand in 3dx1 (Fig. \ref{fig:3dx1}) is much smaller than the ligand in 2zcq (Fig. \ref{fig:2zcq}). Small ligands have fewer opportunities for pairwise contacts with their protein binding partners than larger ligands. The limited number of contacts leads to a weaker $\Psi^\mathrm{bind,np}$, such that $\Xi^\mathrm{pol}$ can play a larger role.

The relative importance of ligand polarization in small ligands may explain the poor performance of binding free energy methods based on a fixed-charge force field in distinguishing molecules that are active and inactive against T4 lysozyme L99A \cite{Xie2017}. In this protein, the L99A mutation forms a pocket known to bind a number of small hydrophobic compounds. \citet{Xie2017} performed binding free energy calculations for a library of 141 small hydrophobic compounds whose thermal activity against T4 lysozyme L99A had been measured. Many of the compounds contained highly polarizable aromatic groups. The best-performing method in \citet{Xie2017} had an area under the receiver operating characteristic curve of 0.74 (0.04) out of 1 for a perfect binary classifier. Binary classification performance could potentially be improved by incorporating the ligand polarization, as described in the current paper.

\subsection{Solvation and polarization can be key drivers of native complex formation}

\begin{figure}
\includegraphics[width=0.8\linewidth]{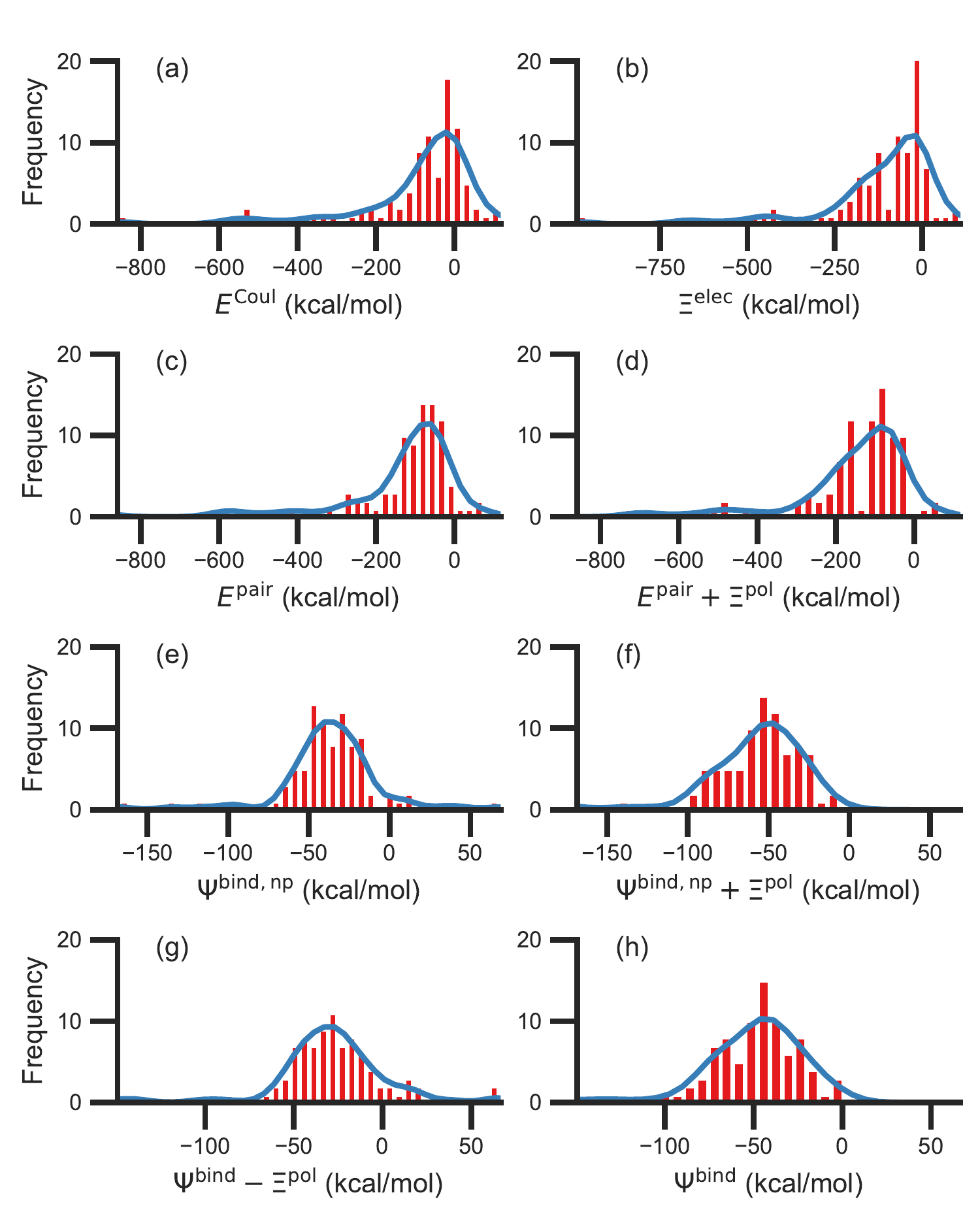}
    \caption{Histograms of intermolecular potential energies: (a) the permanent Coulomb interaction ($E^\mathrm{Coul}$), (b) the electrostatic interaction ($\Xi^\mathrm{elec} = E^\mathrm{Coul} + \Xi^\mathrm{pol}$), (c) the intermolecular pairwise potential energy ($E^\mathrm{pair} = E^\mathrm{vdW} + E^\mathrm{Coul}$), and (d) the intermolecular pairwise potential energy with the polarization energy of the ligand ($E^\mathrm{pair} + \Xi^\mathrm{pol}$) in the gas phase. 
	Histograms of binding energies: the binding energy (e) without considering ligand polarization at all, $\Psi^\mathrm{bind, np}$, and (f) considering ligand polarization for electrostatic interactions but not in the solvation free energy, $\Psi^\mathrm{bind, np} + \Xi^\mathrm{pol}$, (g) considering ligand polarization in the solvation free energy but not for electrostatic interactions, $\Psi^\mathrm{bind} - \Xi^\mathrm{pol}$, or (h) considering ligand polarization both in the electrostatic interactions and the solvation free energy. A similar plot that only considers systems for which $-50 < \Xi^\mathrm{pol} < 0$ kcal/mol is available as Fig. S12 in the Supporting Information. 
	\label{fig:hist_interact}}
\end{figure}

For a number of native complexes, both polarization and solvation were required to compute negative binding energies (Fig. \ref{fig:hist_interact}). Due to the harmonic restraint maintained during minimization, our models closely resemble their native crystal structures. In order for these protein-ligand complexes to adopt these structures, they should have a negative binding energy (presuming that binding results in entropy loss). If calculated binding energies for experimentally observed structures are positive, it suggests a structural modeling issue (e.g. protonation state) or that critical phenomena are not properly described. Intriguingly, the Coulomb interaction energy is positive in a significant fraction of these systems (Fig. \ref{fig:hist_interact}a). Incorporating van der Waals interactions in $E^\mathrm{pair}$ slightly reduces the number of systems in which the interaction energy is positive (Fig. \ref{fig:hist_interact}c). However, these pairwise terms, which are standard to molecular docking, are insufficient to accurately describe all the native complexes with a negative interaction energy. Incorporating a ligand polarization term (Fig. \ref{fig:hist_interact}b \& \ref{fig:hist_interact}d) or solvation energy term (Fig. \ref{fig:hist_interact}e \& g) alone is also insufficient. However, when both polarization and solvation are considered, all the native complexes have a negative binding energy (Fig. \ref{fig:hist_interact}f \& h). Considering both polarization and solvation terms also appears to attenuate the broad range of binding energies observed in $E^\mathrm{pair}$, $E^\mathrm{pair} + \Xi^\mathrm{pol}$, and $\Psi^\mathrm{bind, np}$ (Fig. \ref{fig:hist_interact}f \& h). Using the partial charges $q_A^\mathrm{QM}$ opposed to $q_A^\mathrm{QM}:Q_I$ does not have a qualitative effect on these trends. The trends also hold for systems within the normal range of $-50 < \Xi^\mathrm{pol} < 0$ kcal/mol (Fig. S12 in the Supporting Information). The importance of including ligand polarization and solvation was previously noted by \citet{Kim2016}, who achieved superior performance at binding pose prediction using a protocol that combined atomic charges from QM/MM with solvation compared to using either by themselves.

In cases where the pairwise interaction energy is positive, the change in solvation free energy upon binding can still be negative. Although the electrostatic component of the solvation energy change should have the same sign as the Coulomb interaction energy, formation of a complex reduces the solvent-accessible surface area. An illustrative example occurs in the complex 1z9g. The ligand in this complex is soluble due to a 3-oxopropanoic acid moiety and other hydrophillic components, but it contains an aromatic ring that is buried upon binding to the protein. The decreased solvent-accessible surface area, especially around the aromatic moiety, reduces ordering of solvent in the vicinity of the solute and thereby leads to an increase in entropy.

The lowest $E^\mathrm{Coul}$ are due to phosphate groups. The lowest $E^\mathrm{Coul}$ is observed in the complex 2zcq. The complex contains two Mg$^{2+}$ in close proximity to a negatively-charged phosphate group (Fig. \ref{fig:2zcq}). The complex 1u1b also has a very low $E^\mathrm{Coul}$. The ligand in 1u1b contains four phosphates (Fig. S13 in the Supporting Information). RESP charges on the phosphorus are around 1.4 $e$ and oxygen charges range from -0.4 to 0.8 $e$, leading to a low $E^\mathrm{Coul}$.

\subsection{Solvation but not polarization improves correlation with experimental binding free energies}

An important goal in protein-ligand modeling is the accurate calculation of binding free energies -- which quantify the strength of noncovalent association -- that are consistent with experimentally observed values. 

For several reasons, the computed binding energy $\Delta G^\mathrm{bind}$ is not expected to completely agree with the experimentally measured binding free energy $\Delta G^\mathrm{bind}$ for complexes in the PDBBind Core Set. These reasons include that:
\begin{itemize}
\item The binding free energy $\Delta G^\mathrm{bind}$ is not rigorously equivalent to $\Psi^\mathrm{bind}$, but is actually an exponential average
over the ensemble of the complex \cite{Gallicchio2010, Menzer2018}. Using $\Psi^\mathrm{bind}$ to model $\Delta G^\mathrm{bind}$ is an approximation that neglects entropy.
\item The binding energy model is not exact. For example, the present model does not explicitly treat polarization of the free ligand by solvent, polarization of the protein by the ligand, and the solvation model does not include explicit water.
\item The PDBBind is a heterogeneous data set in which experimental $\Delta G^\mathrm{bind}$ were determined by various modalities and under different experimental conditions. There may be systematic differences between measured $\Delta G^\mathrm{bind}$ that are not considered in our models.
\item On a related note, experimental conditions used to obtain crystal structures and binding affinity data are different. Crystal structures have packing forces and are generally at a lower temperature.
\end{itemize}
Nonetheless, a comparison between computed interaction energies and experimental binding free energies can be informative.

\begin{figure}
	\includegraphics[width=0.9\linewidth]{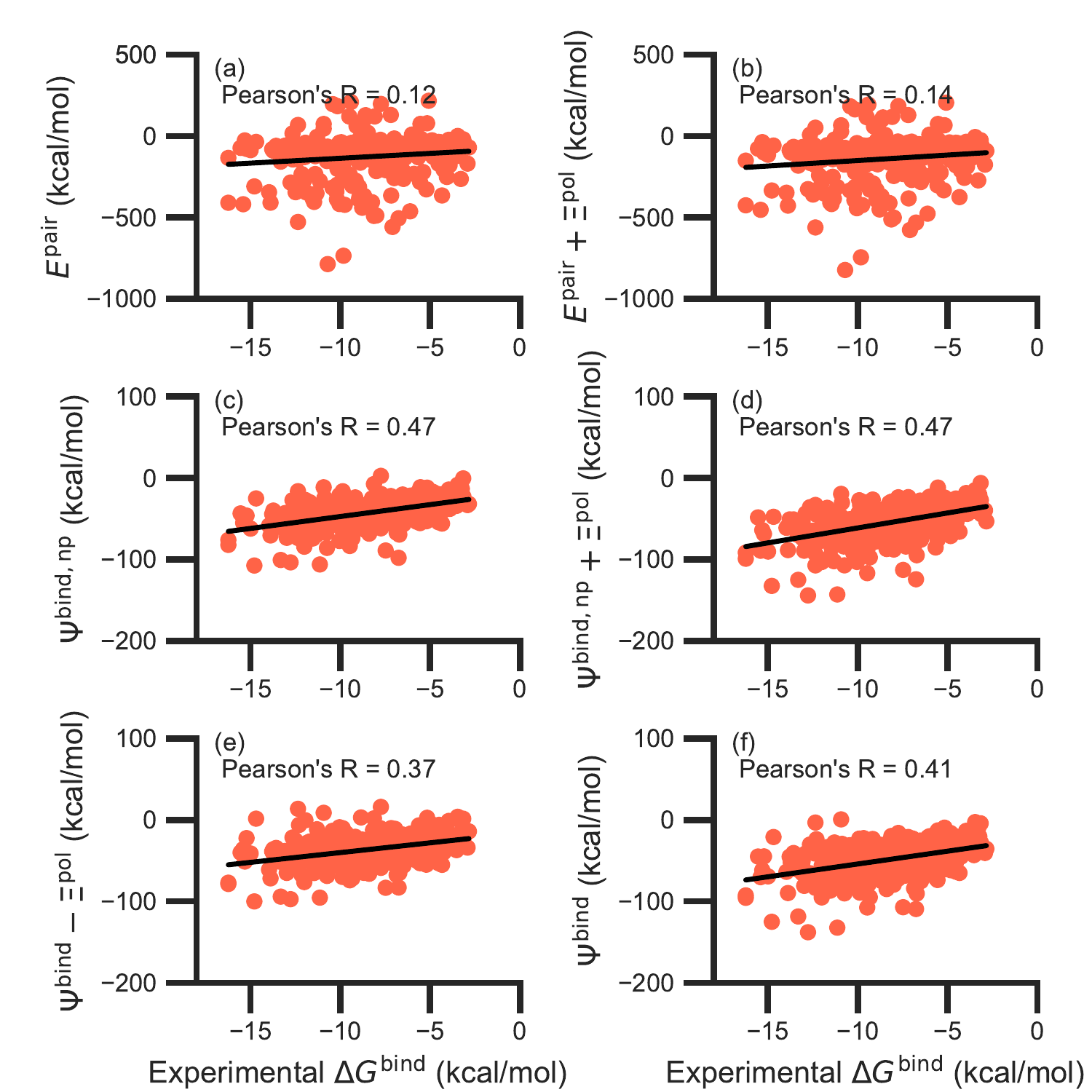}
	\caption{Comparison of interaction energies to experimentally measured binding free energies for complexes with -50 kcal/mol $< \Xi^\mathrm{pol} <$ 0 kcal/mol. Interaction energies are the (a) the intermolecular pairwise potential energy ($E^\mathrm{pair} = E^\mathrm{vdW} + E^\mathrm{Coul}$); (b) the intermolecular pairwise potential energy with the polarization energy of the ligand ($E^\mathrm{pair} + \Xi^\mathrm{pol}$) in the gas phase; the binding energy (c) without considering ligand polarization at all, $\Psi^\mathrm{bind, np}$, and (d) considering ligand polarization for electrostatic interactions but not in the solvation free energy, $\Psi^\mathrm{bind, np} + \Xi^\mathrm{pol}$, (e) considering ligand polarization in the solvation free energy but not for electrostatic interactions, $\Psi^\mathrm{bind} - \Xi^\mathrm{pol}$, or (f) considering ligand polarization both in the electrostatic interactions and the solvation free energy. A similar plot for all complexes is available as Fig. S14 in the Supporting Information.
	\label{fig:scatter_calc_v_exp_interact_subset}}
\end{figure}

While the treatment of solvation is essential, ligand polarization energies have a minimal effect on the correlation between $\Psi^\mathrm{bind}$ and experimental $\Delta G^\mathrm{bind}$ (Fig. \ref{fig:scatter_calc_v_exp_interact_subset} and Fig. S14 in the Supporting Information). If solvation energies are not considered, the distribution of intermolecular pairwise potential energies $E^\mathrm{pair}$ of the protein-ligand complexes is distributed extremely broadly from -1000 kcal/mol to 250 kcal/mol and the correlation between $\Psi^\mathrm{bind}$ and experimental $\Delta G^\mathrm{bind}$ is negligible (Fig. \ref{fig:scatter_calc_v_exp_interact_subset}a\&b and Fig. S14a\&b in the Supporting Information). Incorporating solvation but not polarization significantly improves Pearson's R to 0.47 for complexes where -50 kcal/mol $< \Xi^\mathrm{pol} <$ 0 kcal/mol and 0.40 for complexes where $\Xi^\mathrm{pol} <$ 0 kcal/mol (Fig. \ref{fig:scatter_calc_v_exp_interact_subset}c and Fig. S14c in the Supporting Information). Although the range of computed binding energies is dramatically reduced to -200 kcal/mol to 0 kcal/mol, it is still very broadly distributed compared to the distribution of experimentally measured binding free energies (-16 kcal/mol $< \Delta G^\mathrm{bind} <$ -3 kcal/mol), supporting the idea that a single structure cannot represent an ensemble of structures obtained in experimental conditions. Adding the polarization energy to solvation energies computed without solvation has no effect on the solvation energy (Fig. \ref{fig:scatter_calc_v_exp_interact_subset}d and Fig. S14d in the Supporting Information). In comparison, computing solvation energies using partial charges from the induced dipole diminishes correlation with experiment (Fig. \ref{fig:scatter_calc_v_exp_interact_subset}e\&f and Fig. S14e\&f in the Supporting Information).

In a critical assessment of a number of docking programs and scoring functions across eight different diverse proteins, \citet{Warren2006} concluded that ``no statistically significant relationship existed between docking scores and ligand affinity.'' Our data suggest that the lack of correlation stems from a poor or nonexistent treatment of solvation in the scoring functions. Perhaps due to cancellation of error, neglect of ligand polarization does not appear to be a major factor in the poor performance of docking scores.

\section{Conclusions}

Using a QM/MM approach.\cite{Field1990,Gao1992,Gao1996,Hensen2004}, we computed polarization energies $\Xi^\mathrm{pol}$ for 286 complexes in the PDBBind Core Set \cite{Liu2017c}. The distribution of $\Xi^\mathrm{pol}$, $\Xi^\mathrm{dist}$, and $\Xi^\mathrm{stab}$ were found to be broad and skewed. For properly prepared systems without atoms in unrealistically close contact, these terms all have the expected sign of $\Xi^\mathrm{pol} < 0$, $\Xi^\mathrm{dist} > 0$, and $\Xi^\mathrm{stab} < 0$. The lowest $\Xi^\mathrm{pol}$ were observed in systems where cations are close to ligand atoms. In these systems, the extent of polarization is likely to be overestimated. The importance of including embedding field charges on $\Xi^\mathrm{pol}$ appears to diminish, on average, as an inverse square law. There is no clear relationship between $\Xi^\mathrm{pol}$ and the percentage of highly charged atoms in a protein and molecular polarizability scalar. There is a weak correlation between $\Xi^\mathrm{pol}$ and the Coulomb energy $E^\mathrm{Coul}$. On the other hand, there is a stronger linear correlation between $\Xi^\mathrm{pol}$ and the magnitude of the electric field, the magnitude of the induced dipole moment, and the classical polarization energy. The ligand polarization energy $\Xi^\mathrm{pol}$ is observed to a substantial and system-dependent fraction of the electronic interaction energy and the total interaction energy. In some systems, consideration of ligand polarization and solvation are both essential for calculating negative interaction energies for crystallographic complexes. While consideration of solvation is essential for achieving moderate correlation between interaction energies and experiment, 
we did not observe that the ligand polarization energy $\Xi^\mathrm{pol}$ improves the correlation between the binding energy and experimental binding free energies.

\acknowledgement


We thank Pengyu Ren for the suggestion to compare polarization energies with molecular polarizability.

We thank OpenEye Scientific Software, Inc. for providing academic licenses to their software. This research was supported by the National Institutes of Health (R01GM127712).

\suppinfo

\begin{itemize}
	\item Fig. S1. Histograms of the ligand polarization, distortion, and stabilization energies in the PDBBind Core Set for systems with and without cations.
	\item Fig. S2. Normalized histogram of partial atomic charges for protein atoms in the data set.
	\item Fig. S3. The polarization energy $\Xi^\mathrm{pol}$ as a function of the percentage of charged atoms, number density of highly charged atoms, and Coulomb energy, $E^\mathrm{Coul}$. 
	\item Fig. S4. Polarizability of the ligand ($\alpha_L$) versus the the number of ligand electrons ($N_\mathrm{elec}$) in ligands from the protein-ligand complexes.
	\item Fig. S5. The ligand polarization energy, $\Xi^\mathrm{pol}$, as a function of the magnitude of the electric field. 
	\item Fig. S6. Comparison of electric field estimates.
	\item Fig. S7. The ligand polarization energy, $\Xi^\mathrm{pol}$, as a function of the magnitude of the induced dipole moment.
	\item Fig. S8 The ligand polarization energy, $\Xi^\mathrm{pol}$, as a function of the classical polarization energy.
	\item Fig. S9. The structure of the complex 3tsk of human matrix metalloprotease-12 (MMP12) in complex with L-glutamate motif inhibitor.
	\item Fig. S10. Schematic of protein-ligand complexes in which the ligand center of mass is inside the ligand or the protein.
	\item Fig. S11. Histograms of ratio of the polarization energy of the ligand to intermolecular potential energies and binding energies in all complexes where $\Xi^\mathrm{pol} <$ 0 kcal/mol.
	\item Fig. S12. Histograms of intermolecular potential energies and binding energies in systems where -50 kcal/mol $< \Xi^\mathrm{pol} <$ 0 kcal/mol.
	\item Fig. S13. The structure of the complex 1u1b of bovine pancreatic Ribonuclease A with a ligand.
	\item Fig. S14. Comparison of interaction energies to experimentally measured binding free energies for all complexes where $\Xi^\mathrm{pol} <$ 0 kcal/mol.
	\item Tab. S1. Dependence of the ligand polarization energy on the QM region.
	\item Tab. S2. Complexes with ratios outside of the range of Fig. 6.
\end{itemize}

\bibliography{pdbbind_pol}

\makeatletter
\setcounter{figure}{0} 
\renewcommand{\thefigure}{S\@arabic\c@figure}
\makeatother
\makeatletter
\setcounter{table}{0} 
\renewcommand{\thetable}{S\@arabic\c@table}
\makeatother

\clearpage


\section*{Supporting Information: Histograms}

\begin{figure}
	\includegraphics[width=0.475\linewidth]{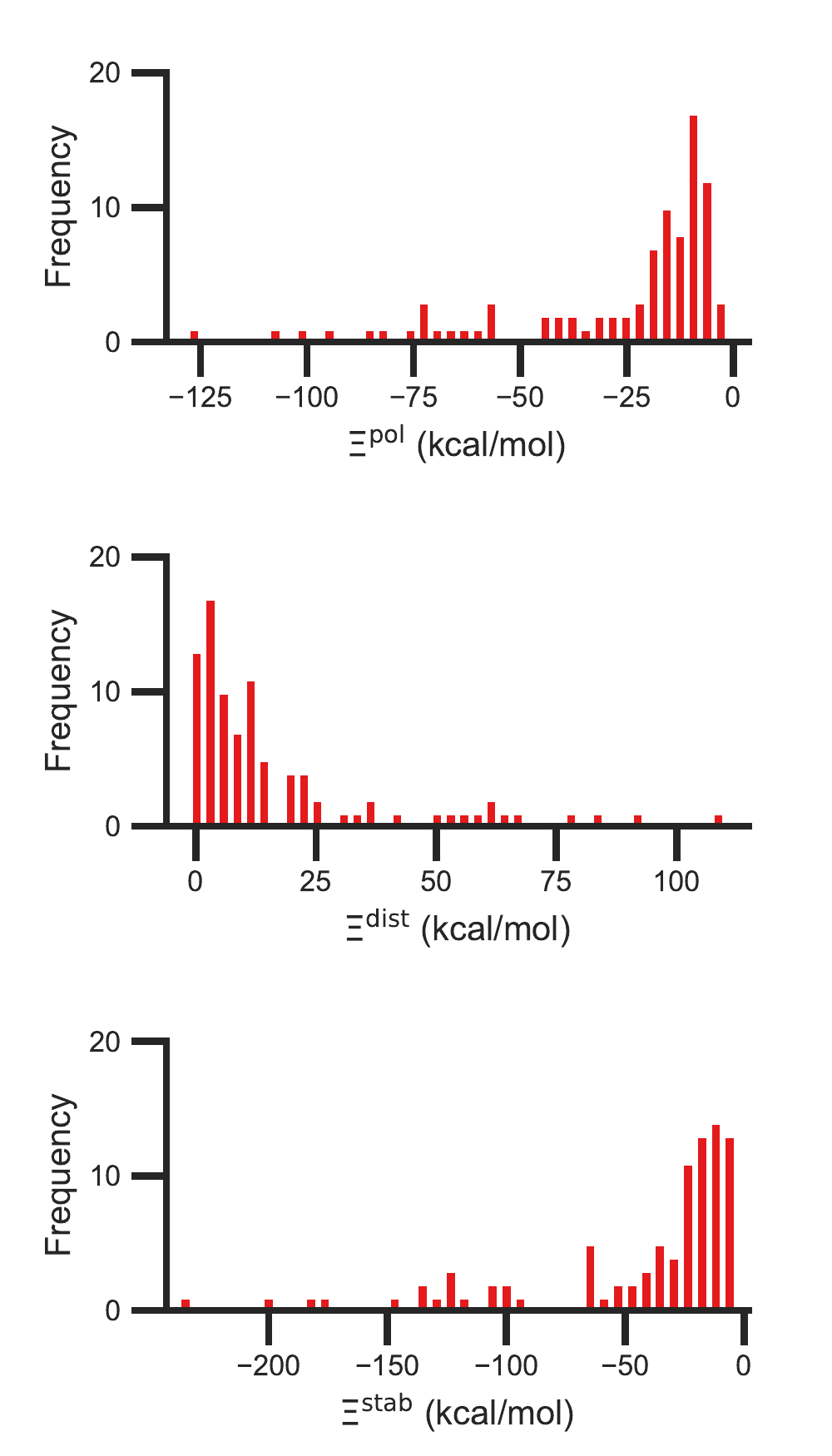}
	\includegraphics[width=0.475\linewidth]{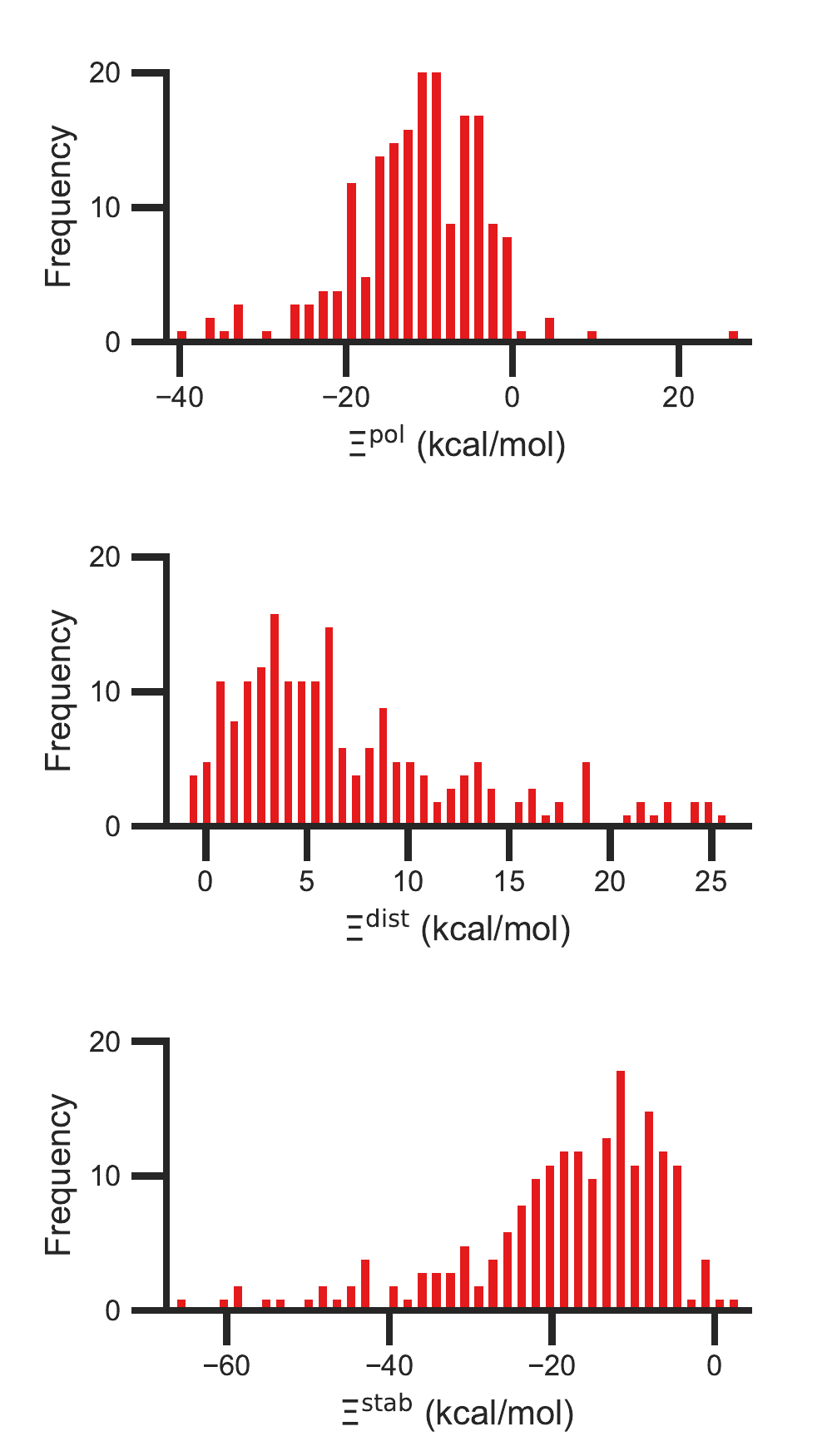}
	\caption{\label{sup_fig:epol}
	Histograms of the ligand polarization (top, $\Xi^\mathrm{pol}$), distortion (middle, $\Xi^\mathrm{dist}$), and stabilization (bottom, $\Xi^\mathrm{stab}$) energies in the PDBBind Core Set for systems with (left) and without (right) cations. The three quantities are related by $\Xi^\mathrm{pol} = \Xi^\mathrm{dist} + \Xi^\mathrm{stab}$.}
\end{figure}

\clearpage

\begin{figure}
    \includegraphics[width=\linewidth] {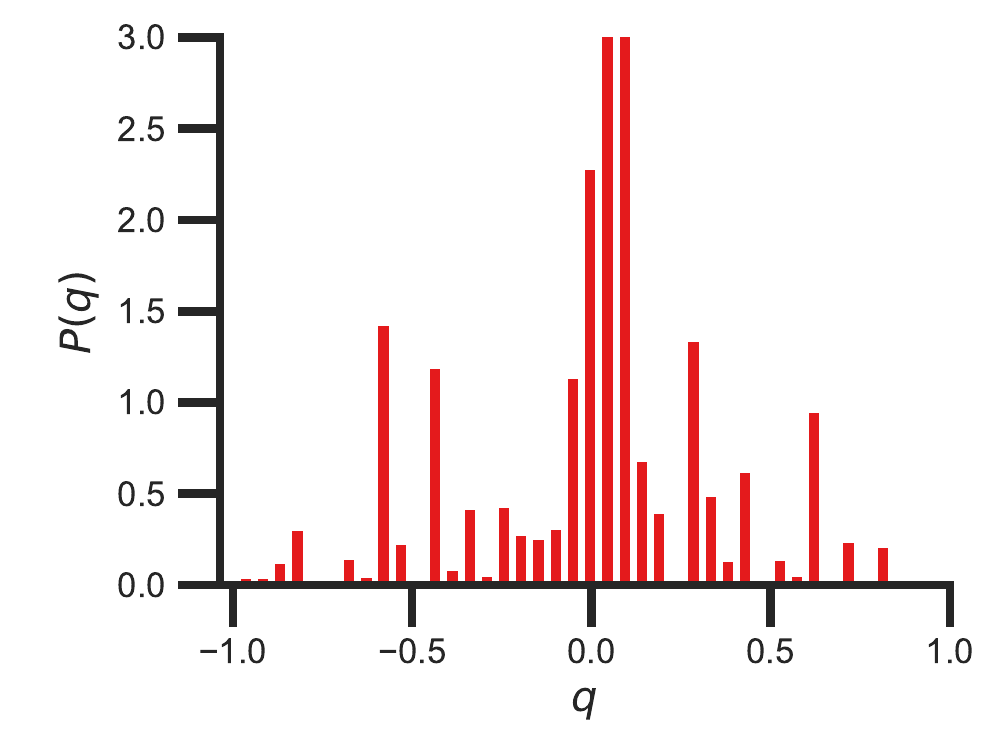}
	\caption{Normalized histogram of partial atomic charges for protein atoms in the data set.
	}
\end{figure}

\clearpage

\section*{Supporting Information: Correlations}

\begin{figure}
    \includegraphics[width=0.9\linewidth] {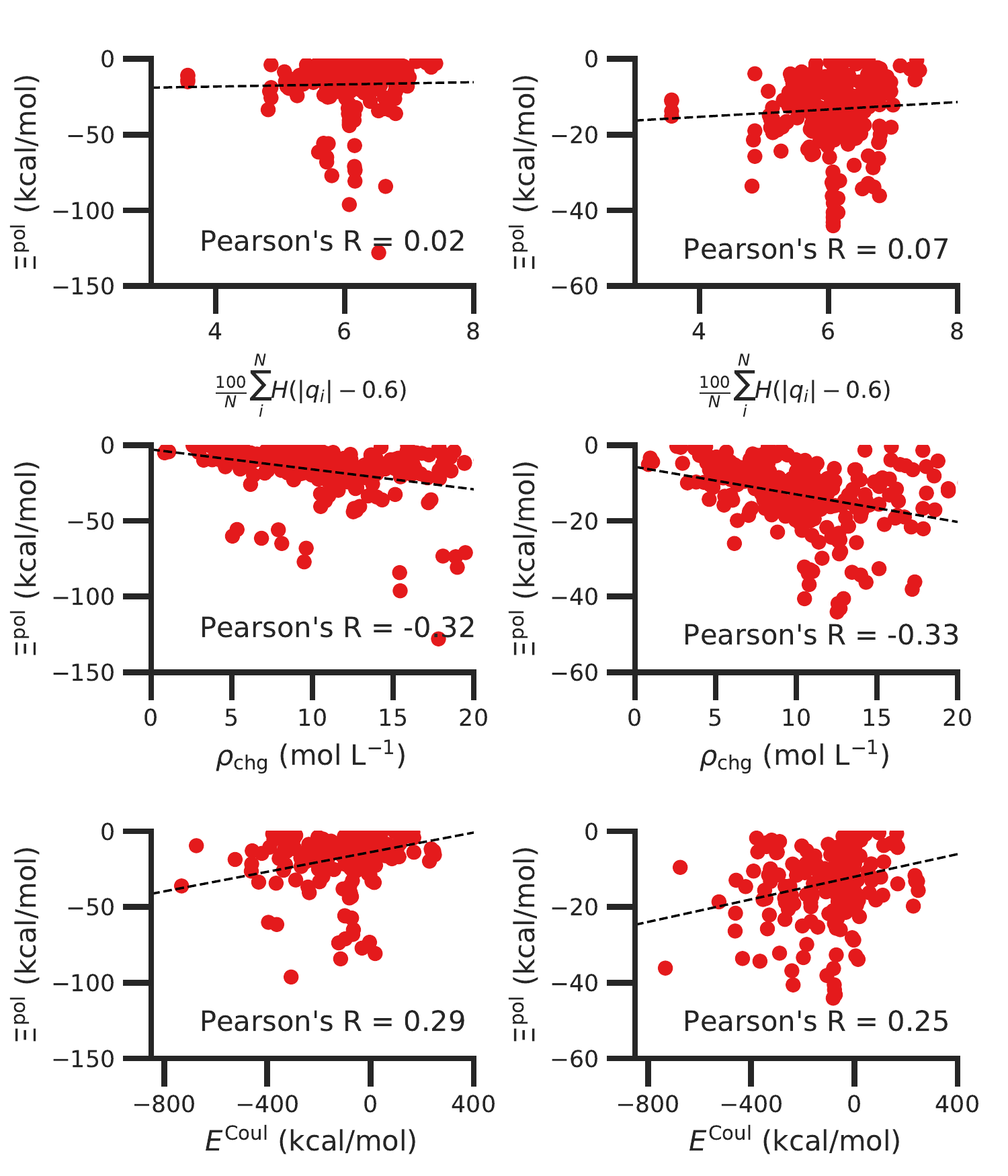}
	\caption{
	The polarization energy $\Xi^\mathrm{pol}$ as a function of the percentage of charged atoms (top), the number density of highly charged atoms (middle), and Coulomb energy $E^\mathrm{Coul}$. Data are included for complexes with $\Xi^\mathrm{pol} <$ 0 kcal/mol (left) or only for complexes with -50 kcal/mol $< \Xi^\mathrm{pol} <$ 0 kcal/mol (right). The number density of charged atoms in a binding site is defined as the number of charged atoms with $|q| > 0.6$ divided by the volume of the site. The volume of the site is the region within 6 \AA ~of any ligand atom.
	\label{sup_fig:scatter_Xipol_3x2prop_rev}
    }
\end{figure}

The percentage of atoms in a protein that are highly charged does not appear to be a significant factor in the ligand polarization energy (Fig. \ref{sup_fig:scatter_Xipol_3x2prop_rev}). In all the systems, only a small percentage of atoms (less than 8\%) have atomic charges of $|q| \geq 0.6$.

The number density of charged atoms is more related with the ligand polarization energy. However, we only observed weak correlation, with Pearson's R being 0.32 for all complexes and 0.33 for complexes for which $\Xi^\mathrm{pol} > -50$ kcal/mol (Fig. \ref{sup_fig:scatter_Xipol_3x2prop_rev}).

It would be reasonable to think that the polarization energy is related to the Coulomb interaction energy. However, the correlation is also weak, with Pearson's R being 0.29 for all complexes and 0.25 for complexes for which $\Xi^\mathrm{pol} > -50$ kcal/mol (Fig. \ref{sup_fig:scatter_Xipol_3x2prop_rev}).

\clearpage

\begin{figure} 
	\includegraphics[width=0.75\linewidth] {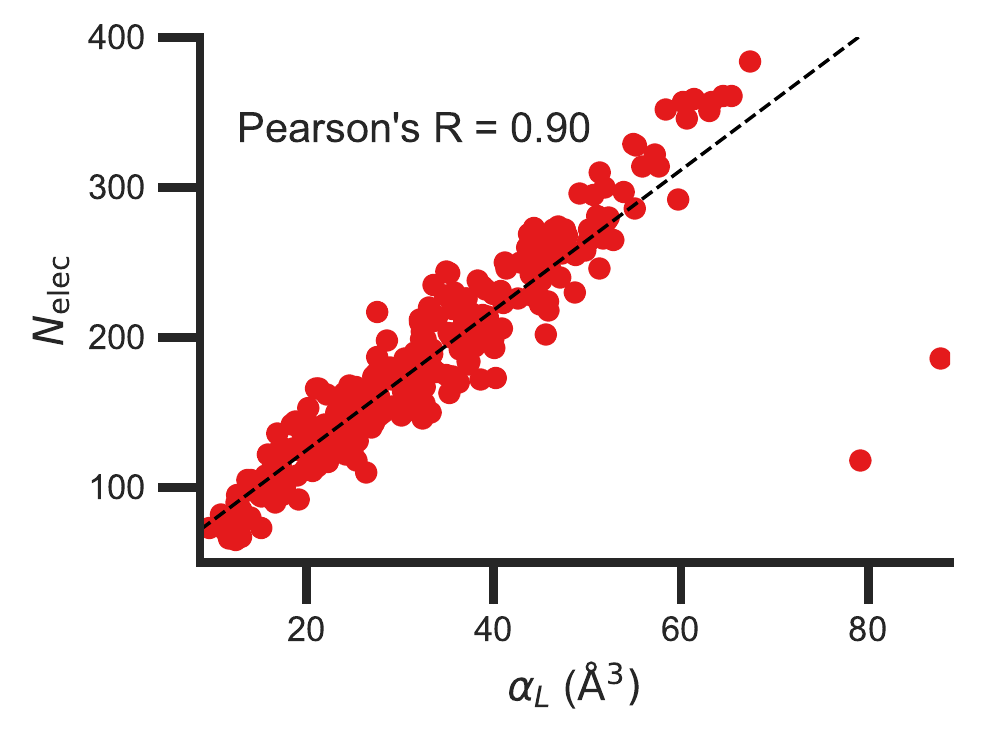}
	\caption{\label{sup_fig:pol_znum}
	Polarizability of the ligand ($\alpha_L$) versus the the number of ligand electrons ($N_\mathrm{elec}$) in ligands from the protein-ligand complexes.
	}
\end{figure}

The molecular polarizability scalar of ligand molecules ($\alpha_L$) has a strong linear correlation with the number of electrons in the system (Fig. \ref{sup_fig:pol_znum}). This observation is reminiscent of one of the properties of halide anions, whose polarizabilities are observed in the following order: F$^- <$ Cl$^- <$ Br$^- <$ I$^-$. However, there is no clear relationship between the molecular polarizability scalar and the ligand polarization energy.

\clearpage 

\begin{figure} 
\includegraphics[width=0.9\linewidth] {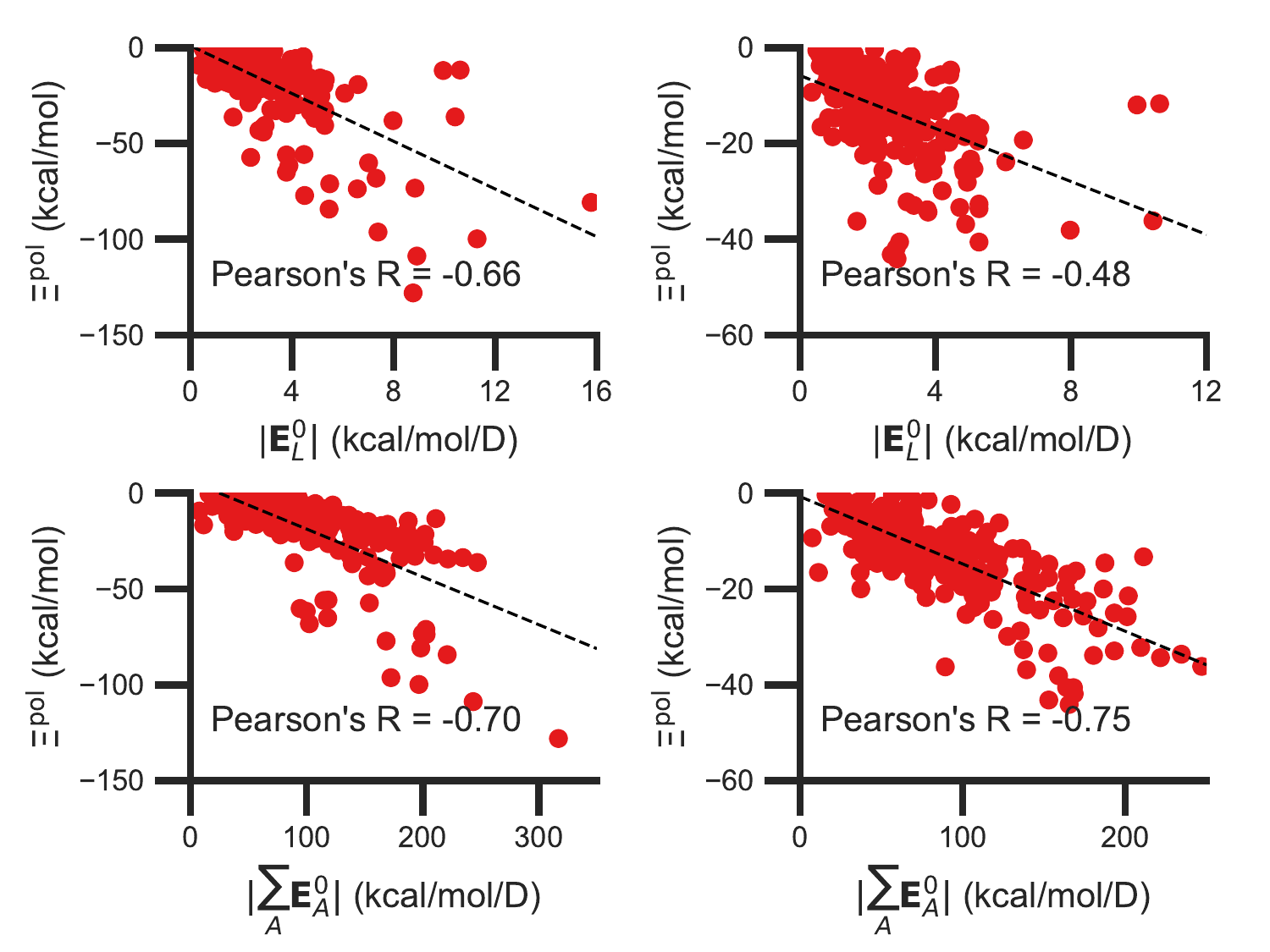}
\caption{The ligand polarization energy, $\Xi^\mathrm{pol}$, as a function of the magnitude of the electric field. 
The electric field vector is either on the ligand center of mass, $|\mathbf{E}_L^0|$, where $\mathbf{E}_L^0$ is from Eq. 22 (top) or the sum of vectors on the ligand atom sites, $\left|\sum_{A \in L} \mathbf{E}_{A}^0 \right|$, where $\mathbf{E}_{A}^0$ is from Eq. 29 (bottom). 
The range of $\Xi^\mathrm{pol}$ is either $\Xi^\mathrm{pol} < 0$ kcal/mol (left) or -50 kcal/mol $< \Xi^\mathrm{pol} < 0$ kcal/mol (right).
	\label{sup_fig:scatter_Xipol_efield}}
\end{figure}

\begin{figure}
    \includegraphics[width=0.75\linewidth] {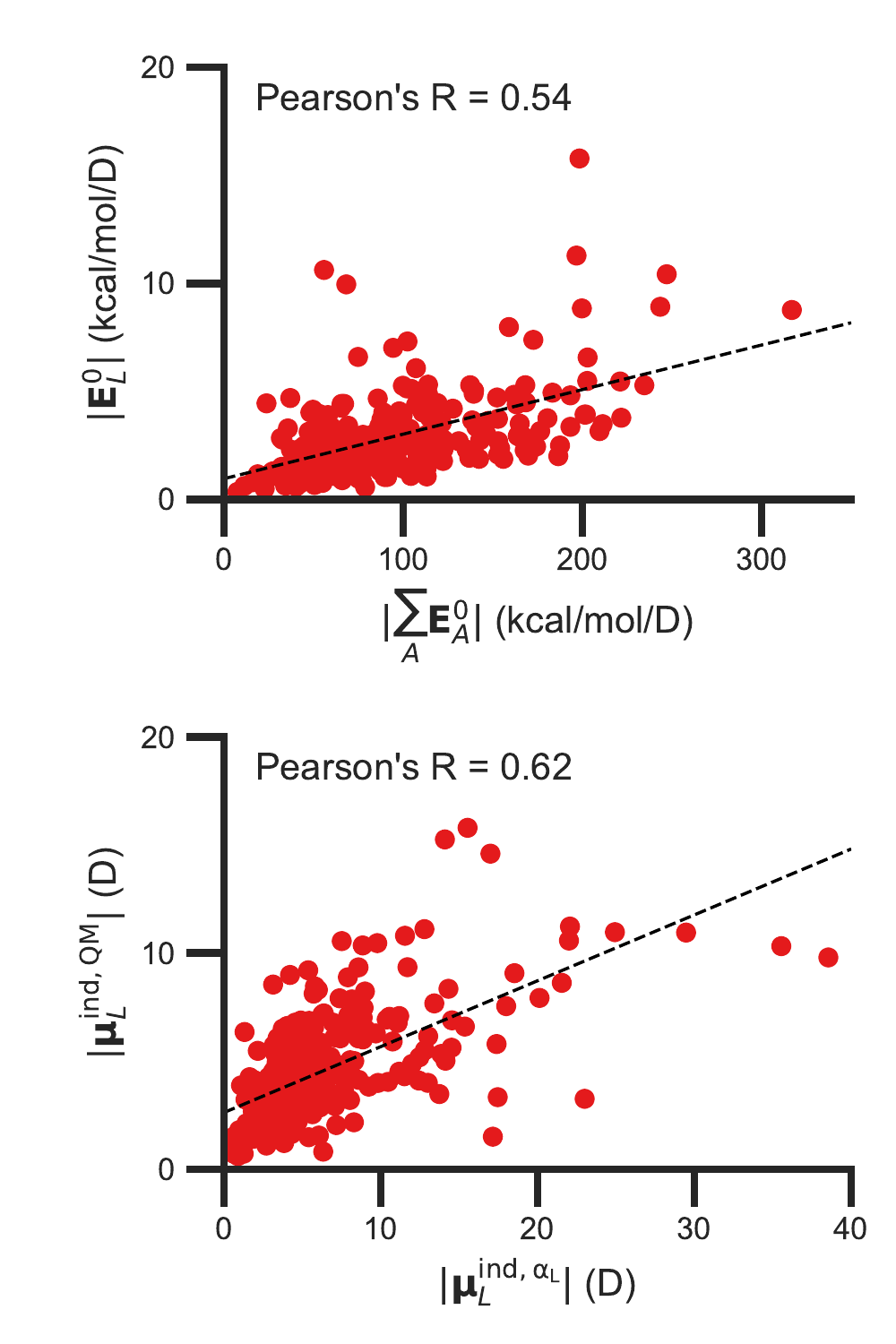}
	\caption{\label{sup_fig:scatter_property_estimates}
    \textbf{Comparison of electric field estimates.} The magnitude of the electric field on the ligand, $|\mathbf{E}_L^0|$, versus of the vector sum of the electric field on all ligand atoms, $|\sum_{A \in L} \mathbf{E}_A^0|$ (top). The magnitude of the induced dipole based on the molecular polarizability tensor, $|\mathbf{\mu}_L^\mathrm{ind, \alpha_L}|$ versus of the induced dipole based on the dipole operator, $|\mathbf{\mu}_L^\mathrm{ind, QM}|$ (bottom).
    }
\end{figure}

In contrast with the aforementioned properties, there is a much clearer relationship between the ligand polarization energy, $\Xi^\mathrm{pol}$, and the magnitude of the electric field (Fig. \ref{sup_fig:scatter_Xipol_efield}). The linear correlation is strong with the magnitude of the electric field on the ligand center of mass, $|\mathbf{E}_{L}^0|$, and even stronger with the magnitude of the total electric field vector active on all ligand atoms, $\left|\sum_{A \in L} \mathbf{E}_{A}^0 \right|$. Intriguingly, in both cases, there appear to be two distinct trends relating the electric field to the magnitude of the electric field; a linear correlation exists in systems where $\Xi^\mathrm{pol} < $ -50 kcal/mol, but the slope is distinct from in systems where  -50 kcal/mol $< \Xi^\mathrm{pol} < $ 0 kcal/mol. The two measures of the electric field are also correlated with each other, with a Pearson's R of 0.54 (Fig. \ref{sup_fig:scatter_property_estimates}). In general, the magnitude of $\left|\sum_{A \in L} \mathbf{E}_{A}^0 \right|$ is greater than the magnitude of $\Xi^\mathrm{pol}$, suggesting that electric field vectors on individual atoms generally point in a similar direction.

\clearpage

\begin{figure} 
	\includegraphics[width=0.9\linewidth] {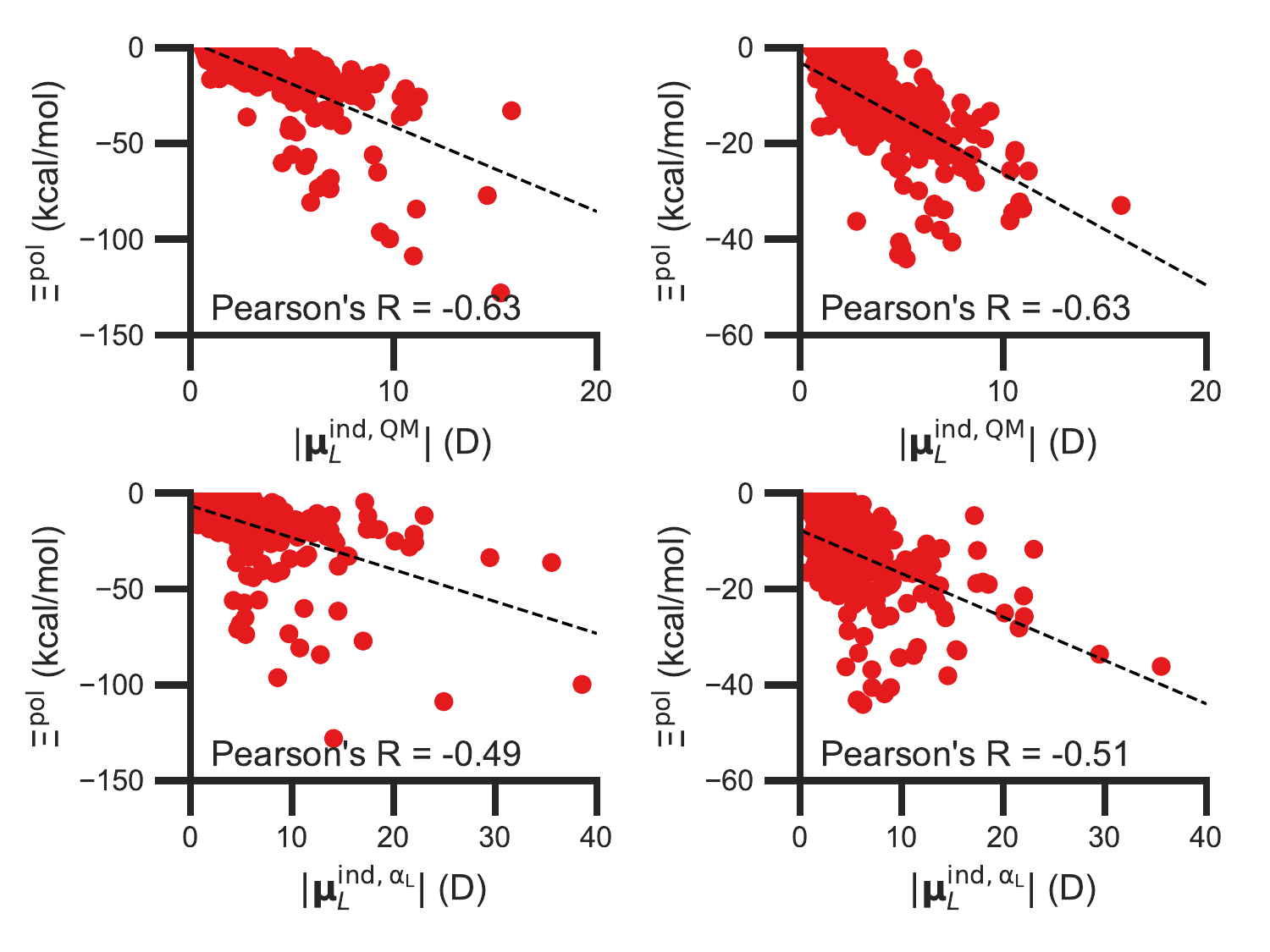}

\caption{The ligand polarization energy, $\Xi^\mathrm{pol}$, as a function of the magnitude of the induced dipole moment. 
The induced dipole moment is either based on wave functions, $|\pmb{\mu}_L^\mathrm{ind, QM}|$, where $\pmb{\mu}_L^\mathrm{ind,\mathrm{QM}}$ is from Eq. 23 (top) or the molecular polarizability tensor, $|\pmb{\mu}_L^\mathrm{ind, \alpha_L}|$, where $\pmb{\mu}_L^{ind, \alpha_L}$ is from Eq. 24 (bottom). 
The range of $\Xi^\mathrm{pol}$ is either $\Xi^\mathrm{pol} < 0$ kcal/mol (left) or -50 kcal/mol $< \Xi^\mathrm{pol} < 0$ kcal/mol (right).
\label{sup_fig:scatter_Xipol_mu}}
\end{figure}

Similarly, the ligand polarization energy $\Xi^\mathrm{pol}$ is also correlated with the magnitude of the induced dipole moment on the ligand. There is a stronger correlation between the ligand polarization energy $\Xi^\mathrm{pol}$ and the induced dipole based on the wave functions $\pmb{\mu}_L^\mathrm{ind, QM}$ (Eq. 23) than the induced dipole based on the molecular polarizability tensor $\pmb{\mu}_L^\mathrm{ind, \alpha_L}$ (Fig. \ref{sup_fig:scatter_Xipol_mu}). The latter quantity, $\pmb{\mu}_L^\mathrm{ind, \alpha_L}$, which is ultimately based on three pairs of point charges, does not perfectly recapitulate polarizability of the more complex embedding field; Pearson's R is 0.62 (Fig. \ref{sup_fig:scatter_property_estimates}.)

\clearpage

\begin{figure} 
\includegraphics[width=0.9\linewidth] {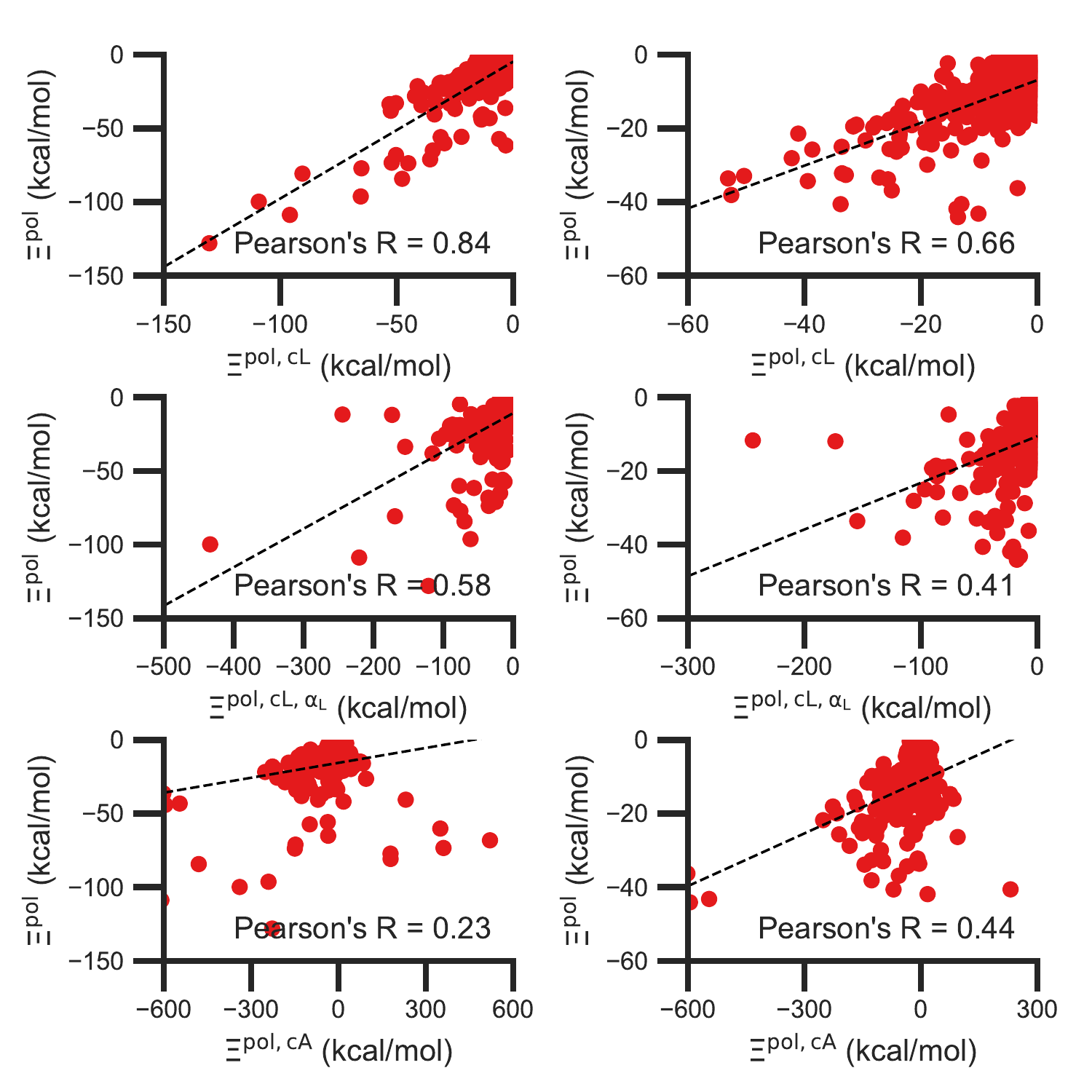}
\caption{The ligand polarization energy, $\Xi^\mathrm{pol}$, as a function of the classical polarization energy.
The classical polarization energy is either $\Xi^\mathrm{pol, cL}$ (Eq. 22), using Eq. 23 for the induced dipole moment (top), $\Xi^\mathrm{pol, cL, \alpha_L}$ (Eq. 22), using Eq. 24 for the induced dipole moment (middle), or $\Xi^\mathrm{pol, cA}$ (Eq. 27). 
The range of $\Xi^\mathrm{pol}$ is either $\Xi^\mathrm{pol} < 0$ kcal/mol (left) or -50 kcal/mol $< \Xi^\mathrm{pol} < 0$ kcal/mol (right).
	\label{sup_fig:scatter_Xipol_QMvClassical}}
\end{figure}

In addition to the strong relationship between the ligand polarization energy $\Xi^\mathrm{pol}$ and both the magnitude of the electric field and the induced dipole, there is also a clear correspondence between the ligand polarization energy $\Xi^\mathrm{pol}$ and the classical polarization energy $\Xi^\mathrm{pol,cL}$ (Fig. \ref{sup_fig:scatter_Xipol_QMvClassical}). The clear correlation between the two quantities suggests that the classical model of the ligand as a single dipole in an electric field is a reasonable explanation for the quantum behavior. In contrast, the correlation between $\Xi^\mathrm{pol}$ and $\Xi^\mathrm{pol,cA}$ is much weaker, which indicates that the classical model of the ligand as a set of atom-centered dipoles is a poor description of the quantum phenomenon. The correlation is stronger between $\Xi^\mathrm{pol}$ and $\Xi^\mathrm{pol,cL}$ than between $\Xi^\mathrm{pol}$ and $\Xi^\mathrm{pol,cL, \alpha_L}$ because the molecular polarizability model does not perfectly capture the induced dipole moment (Fig. \ref{sup_fig:scatter_property_estimates}). 

\clearpage

\section*{Supporting Information: Limitations of molecular polarizability model}

\begin{figure} 
\includegraphics[width=0.4\linewidth] {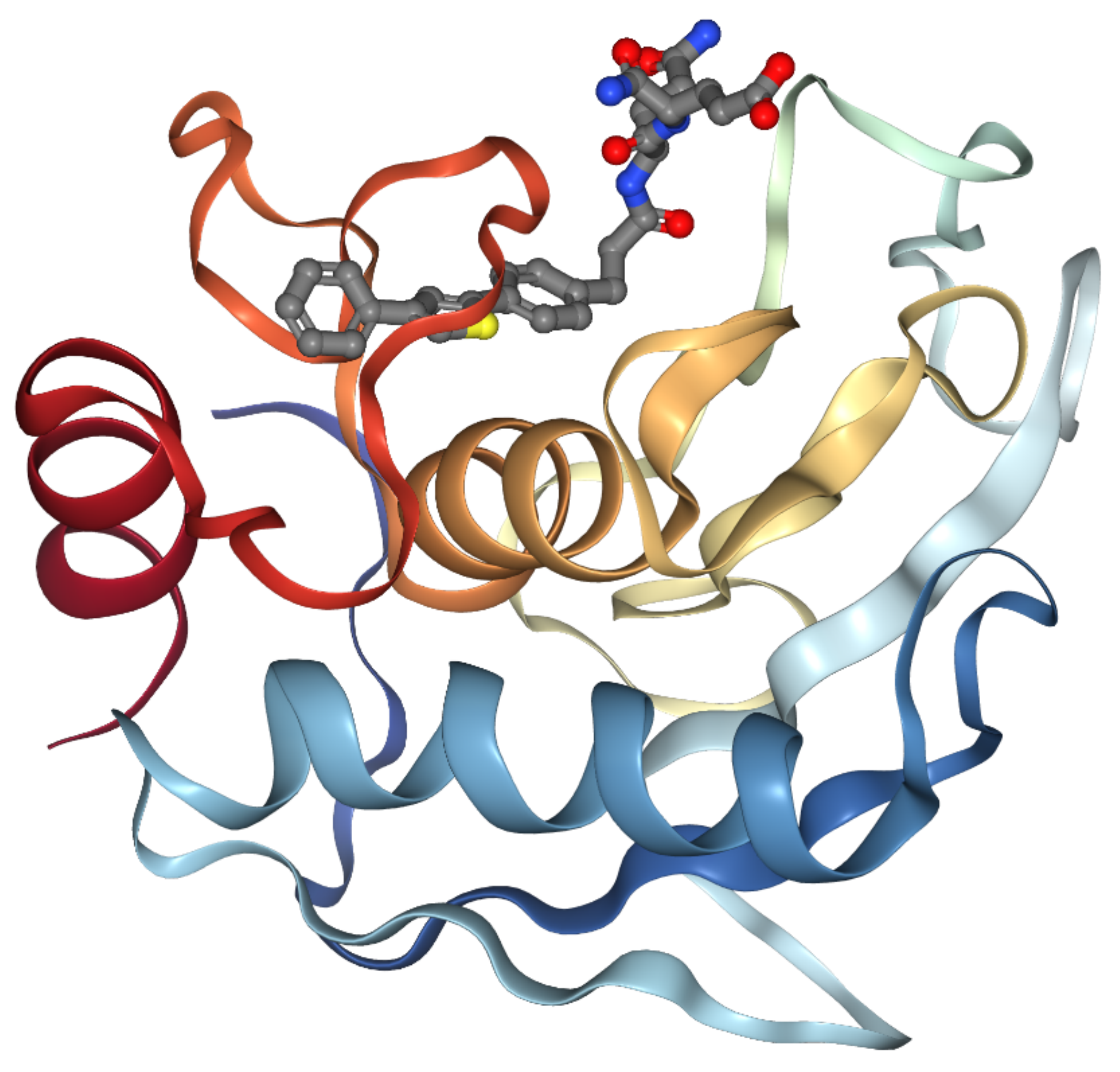}
\caption{
The structure of the complex 3tsk of human matrix metalloprotease-12 (MMP12) in complex with L-glutamate motif inhibitor. The center of mass of the ligand is placed inside the protein.
\label{sup_fig:3tsk}
}
\end{figure}

\begin{figure} 
\includegraphics[width=0.8\linewidth] {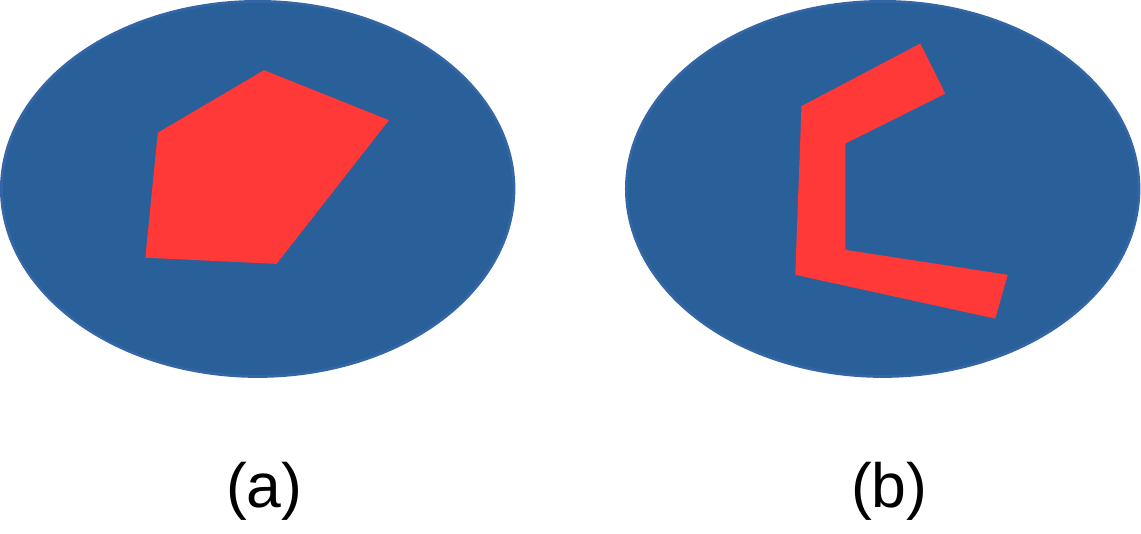}
\caption{Schematic of protein-ligand complexes in which the ligand center of mass is inside the ligand or the protein. The ligand is colored red and protein blue. In (a), the center of mass of the ligand is placed inside the ligand, whereas in (b) the center of mass of the ligand is inside the protein.
\label{sup_fig:protein_ligand}
}
\end{figure}

In many cases, the failure of the molecular polarizability to recapitulate the induced dipole is due to the location of the ligand center of mass. For most complexes, the magnitude of the induced dipole moment based on the molecular polarizability tensor, $|\mathbf{E}_{L\pmb{\mu}_{L}}^{\mathrm{ind}, \alpha_L}|$ is comparable to the magnitude of the induced dipole from the quantum mechanical operator, $|\pmb{\mu}_{L}^\mathrm{ind, QM}|$. However, in nearly 8\% of complexes, $|\mathbf{E}_{L\pmb{\mu}_{L}}^{\mathrm{ind}, \alpha_L}|$ is much larger than $|\pmb{\mu}_{L}^\mathrm{ind, QM}|$. In many of these cases, such as 3tsk (Fig. \ref{sup_fig:3tsk}), the ligand center of mass is within the protein (Fig. \ref{sup_fig:protein_ligand}). Because the ligand center of mass is within the protein, it is very close to embedding field charges and the magnitude of the electric field is particularly strong.

\clearpage

\clearpage

\section*{Supporting Information: Other Figures}

\begin{figure}
	\includegraphics[width=0.45\linewidth]{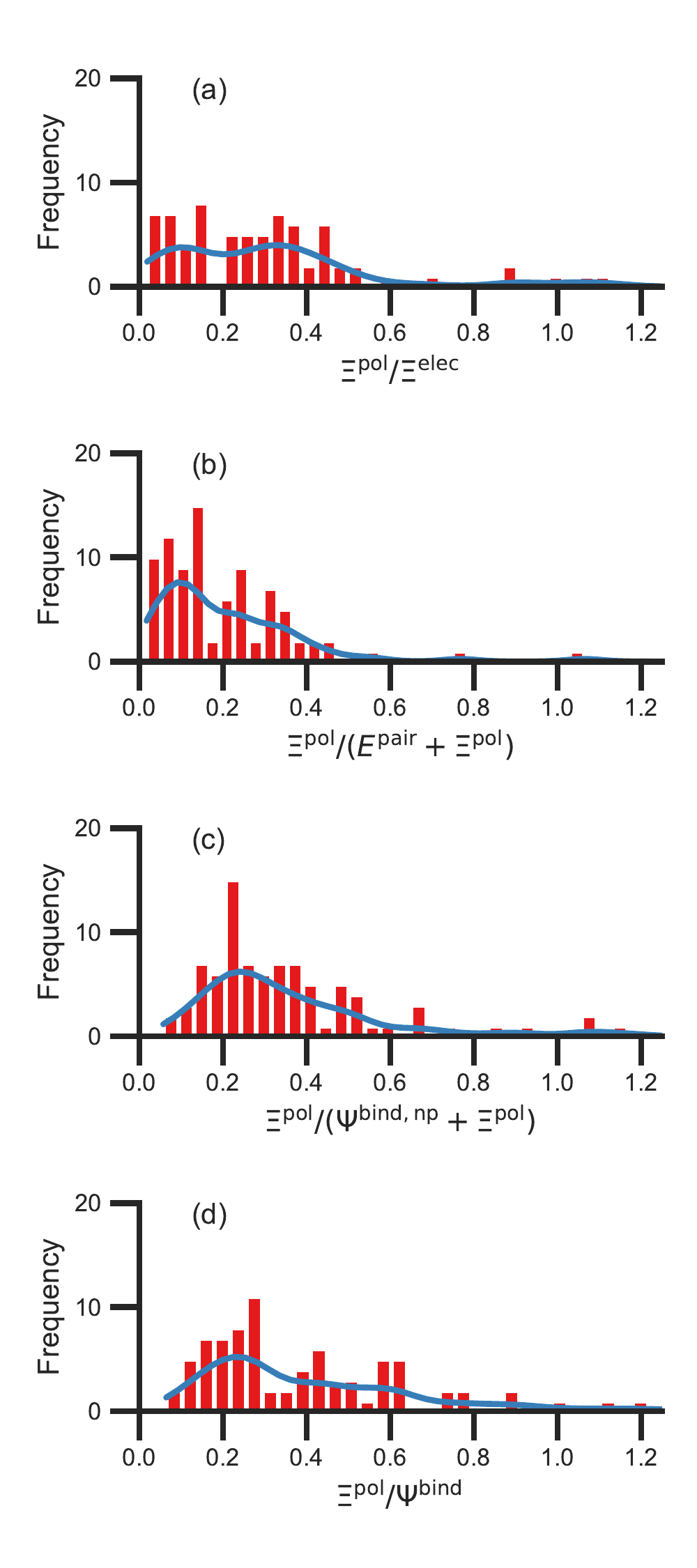}
	\caption{Histograms of ratio of the polarization energy of the ligand to (a) the electrostatic interaction ($\Xi^\mathrm{elec} = E^\mathrm{Coul} +\Xi^\mathrm{pol} $), (b) the intermolecular pairwise potential energy with the ligand polarization energy ($E^\mathrm{pair} + \Xi^\mathrm{pol}$), (c)  the  binding energy without considering ligand polarization in the solvation free energy ($\Psi^\mathrm{bind,np} + \Xi^\mathrm{pol}$), and (d) the binding energy with considering ligand polarization in the solvation free energy ($\Psi^\mathrm{bind}$).
	Data are from all complexes where $\Xi^\mathrm{pol} <$ 0 kcal/mol.
	The histograms are truncated at a ratio of 1.25.
}
\end{figure}

\clearpage

\begin{figure} 
	\includegraphics[width=0.75\linewidth] {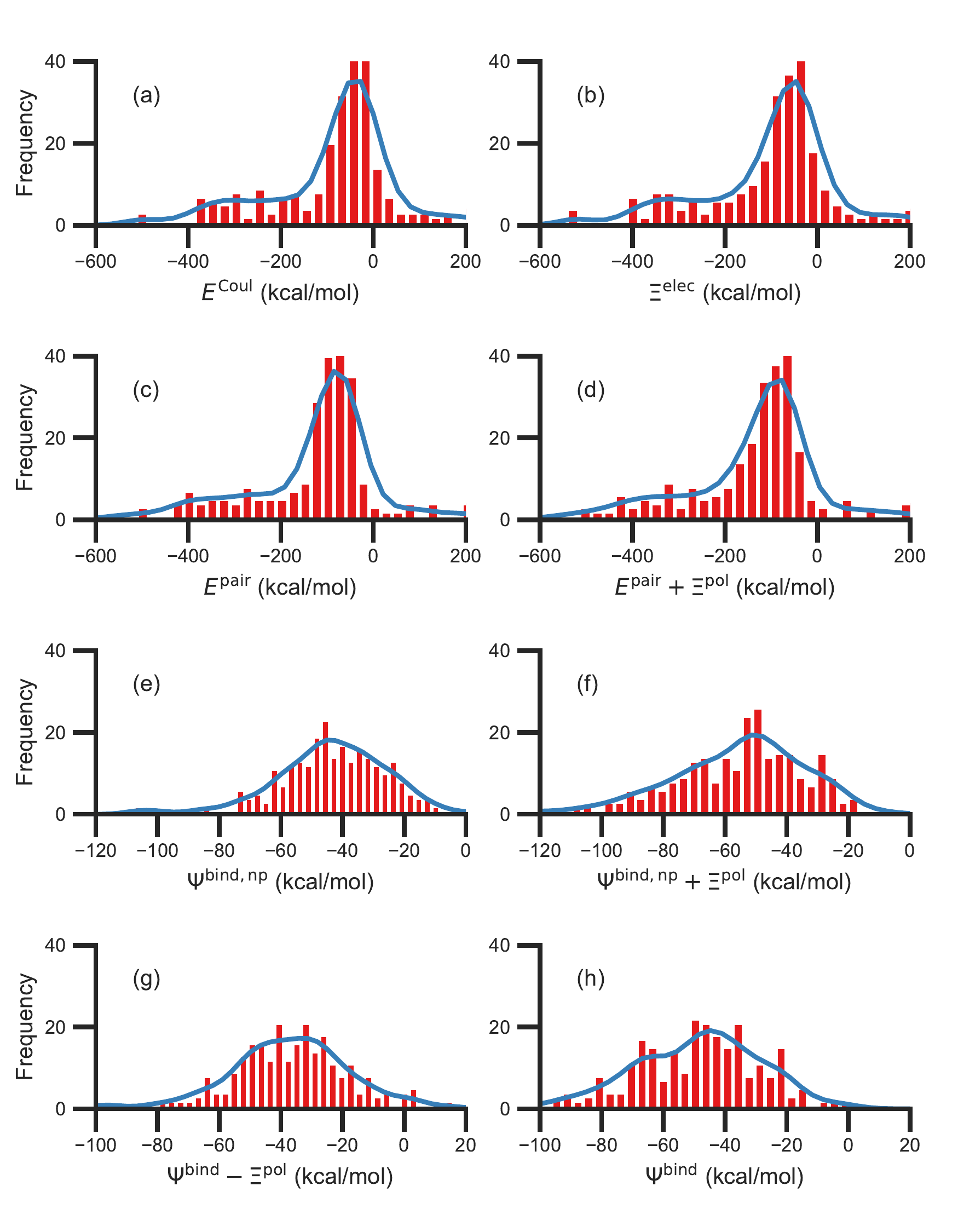}  
    \caption{Histograms of intermolecular potential energies in systems where -50 kcal/mol $< \Xi^\mathrm{pol} <$ 0 kcal/mol: (a) the permanent Coulomb interaction ($E^\mathrm{Coul}$), (b) the electrostatic interaction ($\Xi^\mathrm{elec} = E^\mathrm{Coul} + \Xi^\mathrm{pol}$), (c) the intermolecular pairwise potential energy ($E^\mathrm{pair} = E^\mathrm{vdW} + E^\mathrm{Coul}$), and (d) the intermolecular pairwise potential energy with the polarization energy of the ligand ($E^\mathrm{pair} + \Xi^\mathrm{pol}$) in the gas phase. 
	Histograms of binding energies: the binding energy (e) without considering ligand polarization at all, $\Psi^\mathrm{bind, np}$, and (f) considering ligand polarization for electrostatic interactions but not in the solvation free energy, $\Psi^\mathrm{bind, np} + \Xi^\mathrm{pol}$, (g) considering ligand polarization in the solvation free energy but not for electrostatic interactions, $\Psi^\mathrm{bind} - \Xi^\mathrm{pol}$, or (h) considering ligand polarization both in the electrostatic interactions and the solvation free energy. 
	A similar plot that includes systems containing cations is available in the main text.
	}
\end{figure}

\clearpage

\begin{figure} 
	\centering
	\begin{tabular}{c c}
		\includegraphics[width=0.5\linewidth] {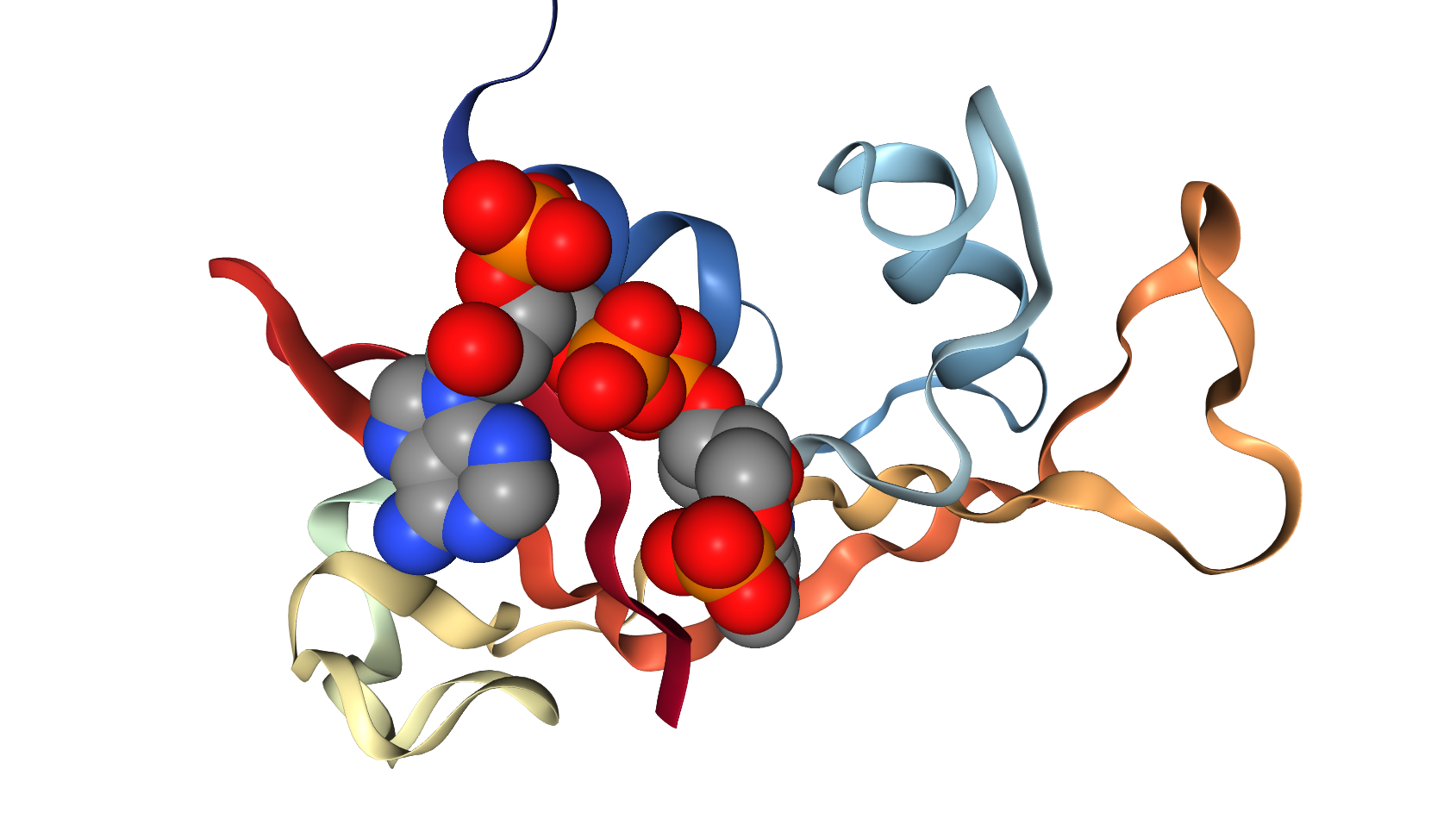} & 
	    \includegraphics[width=0.3\linewidth] {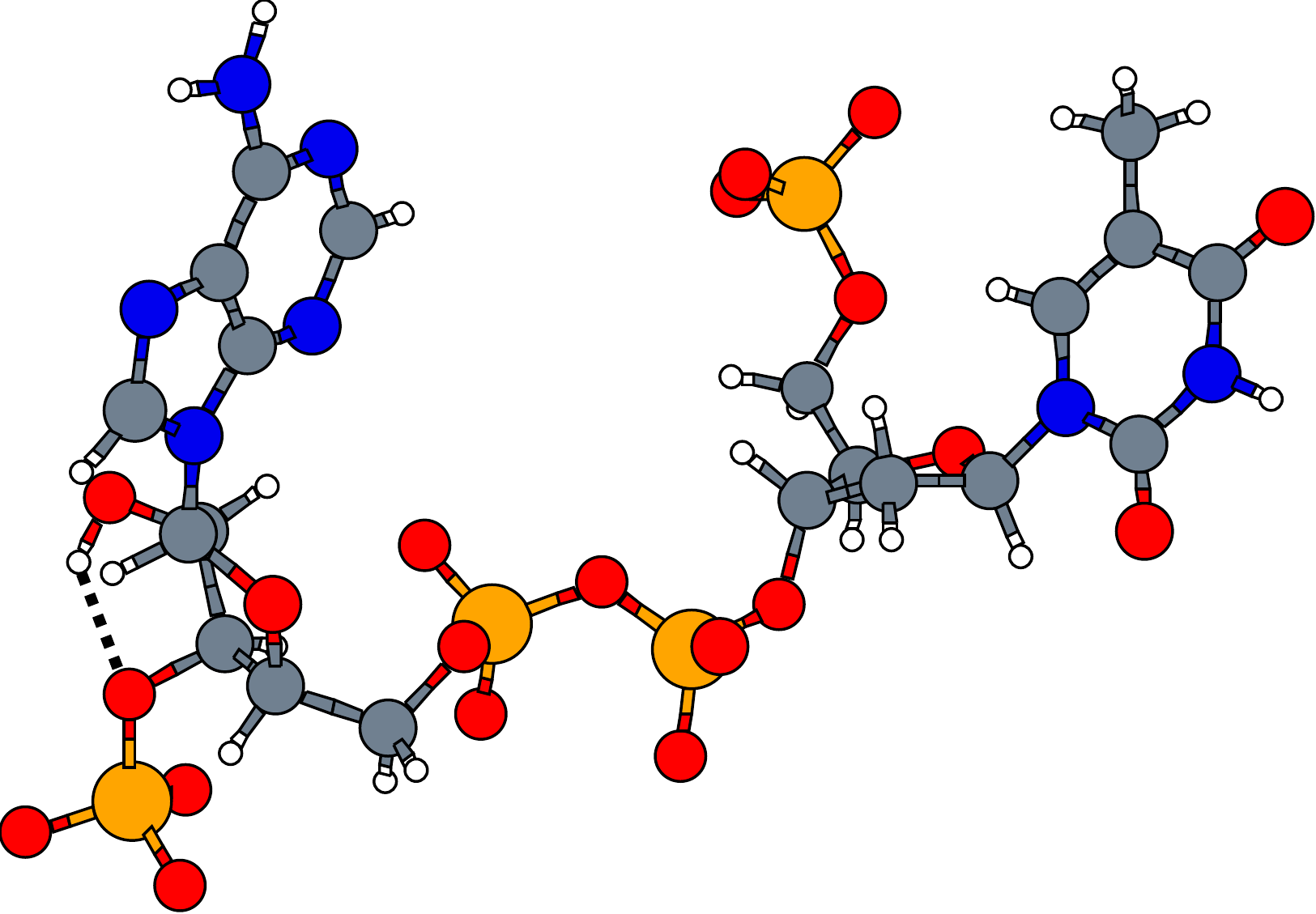} \\
		(a) & (b) \\
	\end{tabular}
	\caption{\label{sup_fig:sup_1u1b}
		(a) The structure of the complex 1u1b of bovine pancreatic Ribonuclease A with the ligand (3$^\prime$-phosphothymidine (3$^\prime$-5$^\prime$)-pyrophosphate adenosine 3$^\prime$-phosphate) and (b) the molecular structure of the ligand.}
\end{figure}

\clearpage


\begin{figure}
	\includegraphics[width=0.9\linewidth]{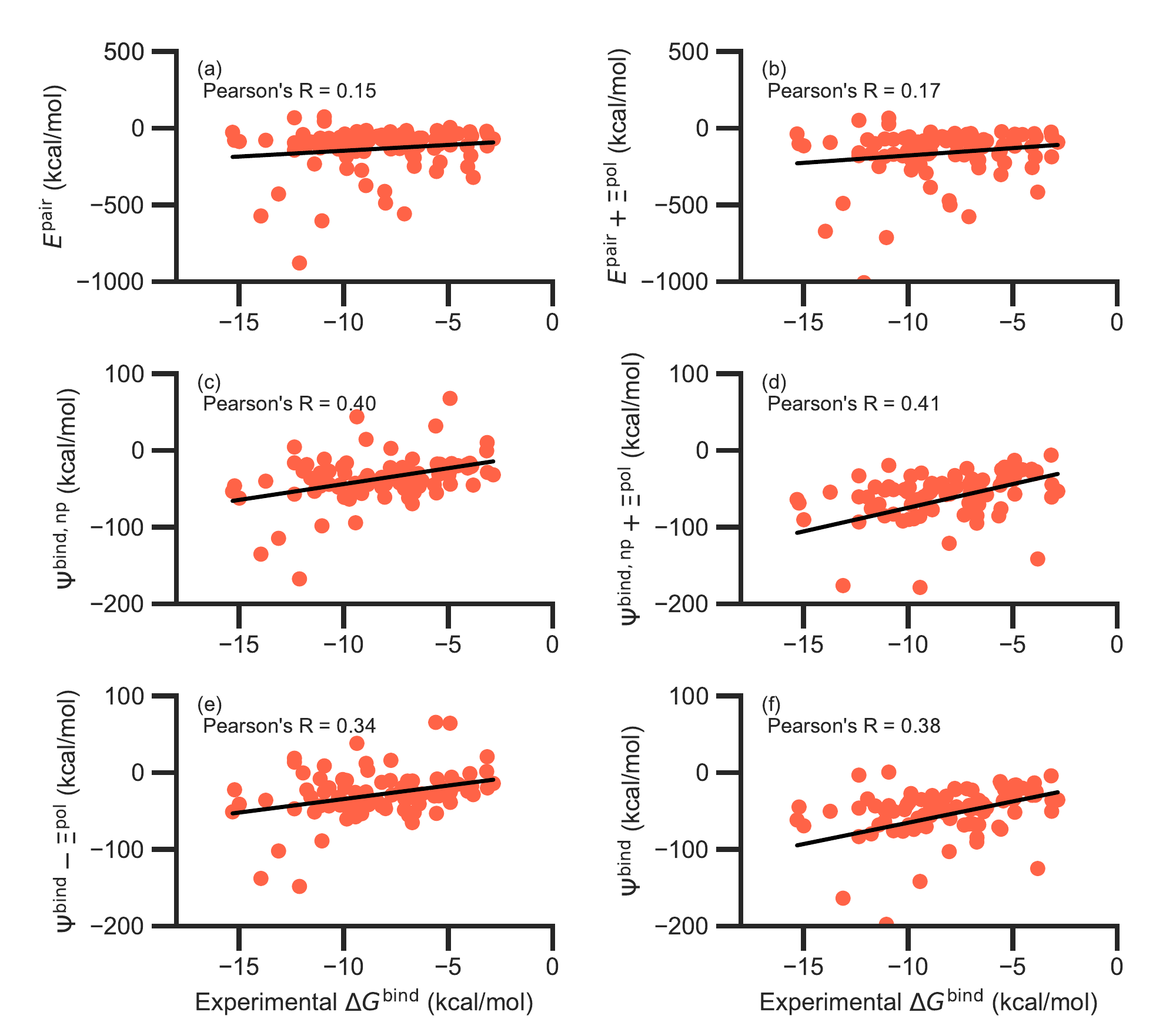}
	\caption{Comparison of interaction energies to experimentally measured binding free energies for all complexes where $\Xi^\mathrm{pol} <$ 0 kcal/mol. Interaction energies are the (a) the intermolecular pairwise potential energy ($E^\mathrm{pair} = E^\mathrm{vdW} + E^\mathrm{Coul}$); (b) the intermolecular pairwise potential energy with the polarization energy of the ligand ($E^\mathrm{pair} + \Xi^\mathrm{pol}$) in the gas phase; the binding energy (c) without considering ligand polarization at all, $\Psi^\mathrm{bind, np}$, and (d) considering ligand polarization for electrostatic interactions but not in the solvation free energy, $\Psi^\mathrm{bind, np} + \Xi^\mathrm{pol}$, (e) considering ligand polarization in the solvation free energy but not for electrostatic interactions, $\Psi^\mathrm{bind} - \Xi^\mathrm{pol}$, or (f) considering ligand polarization both in the electrostatic interactions and the solvation free energy.}
	\label{sup_fig:binding}
\end{figure}

\clearpage

\section*{Supporting Information: Estimated Overpolarization}

When cations are close to ligands, the extent of ligand polarization is likely overestimated by the QM/MM scheme used in this paper. To assess the extent of overpolarization, we performed some calculations in which cations were included in the QM region. In this modified scheme, there is no direct way to isolate the ligand polarization energy from energy of the complex. Instead, we estimated the induced dipole of the ligand based on RESP atomic charges,
\begin{eqnarray}
\mu_L^\mathrm{ind,RESP} = \sum_{A \in L} (q_A^\mathrm{QM:Q_L} - q_A^\mathrm{QM}) \pmb{R}_A.
\end{eqnarray}
The ligand polarization energy is then computed by,
\begin{eqnarray}
\Xi^\mathrm{pol, RESP} &=& 
- \pmb{\mu}_L^\mathrm{ind,RESP} \cdot \mathbf{E}_L^0, 
\end{eqnarray}
where 
$\mathbf{E}_L^0$ is the electric field acting on the center of mass of the ligand.

In the selected systems where cations are very close to ligand atoms, the ligand polarization energy estimated with the main QM/MM scheme in this paper is likely too low (Table \ref{tab:Xi_pol_RESP}).
When the only ligand is in the QM region, the estimated ligand polarization energy is fairly consistent; $\Xi^\mathrm{pol}(L) \sim \Xi^\mathrm{pol,cL} (L) \sim \Xi^\mathrm{pol,RESP} (L)$.
When the QM region is expanded to include cations,
the ligand polarization energy based on RESP atomic charges is significantly higher. 
For 3dx1 and 3dx2, it is about 20 kcal/mol higher. 
For 2zcq, where $\Xi^{\mathrm{pol}}$ is especially low, $\Xi^{\mathrm{pol,RESP}} (LC)$ has the opposite sign!

\begin{table}
	\caption{\label{tab:Xi_pol_RESP} Dependence of the ligand polarization energy on the QM region. $\Xi^\mathrm{pol}(X)$, $\Xi^\mathrm{pol,cL}(X)$, and $E^\mathrm{pol,RESP}(X)$ are from Eqs. 6, 22, and S2, respectively, with the QM region of $X$. Here,  either the ligand ($L$) or the ligand and cations ($LC$) are included in the QM region. 
	The unit of the polarization energy is in kcal/mol.}
\begin{tabular}{| l | r | r | r | r | }
\hline \hline
PDB ID &  ~~$\Xi^\mathrm{pol} (L)$~~ &~~$\Xi^\mathrm{pol,cL} (L)$~~ & $ \Xi^\mathrm{pol,RESP} (L)$~~ & $\Xi^\mathrm{pol,RESP} (LC)$ \\
\hline 
3dx1 & -80.77 & -92.20 & -87.55 & -69.11 \\
3dx2 & -73.34 & -51.71 & -49.51 & -30.77 \\
2zcq & -128.01 &-132.86 & -128.67 & 109.02 \\
\hline \hline
\end{tabular} 

\end{table}

\clearpage

\begin{table}
	\caption{\label{tab:large_ratios} Complexes with ratios outside of the range of Fig. 7.}

		\begin{tabular}{| l | r | r | r |  }
				\hline \hline
			PDB ID & ~~$\Xi^\mathrm{pol}$~~ & $\Xi^\mathrm{elec}$~~ & $\Xi^\mathrm{pol}/\Xi^\mathrm{elec}$~~ \\
			\hline
			1mq6  &   -10.590~~ &      48.843~~ &      -0.217~~ \\
			1o3f    &  -18.158~~    &   56.933~~   &     -0.319~~  \\
			1oyt    &  -18.273~~    &   -3.716~~     &   4.917~~ \\
			3gc5   &   -28.756~~   &   -12.884~~   &     2.232~~  \\
			3ge7   &   -33.854~~    &    2.873~~    &  -11.785~~  \\
			3gy4   &   -19.494~~   &   -12.887~~    &    1.513~~ \\
			3ui7    &  -16.902~~     &  93.232~~    &   -0.181~~  \\
			3uuo   &    -8.104~~  &    108.339~~     &  -0.075~~  \\
			4abg    &  -12.907~~  &     -4.975~~     &   2.594~~  \\
			4ea2   &   -32.932~~   &    21.277~~    &   -1.548~~  \\
			4llx     & -12.174~~     & -6.190~~      &  1.967~~  \\
			4mme  &    -10.732~~  &     21.803~~  &     -0.492~~  \\
			4msc    &  -18.087~~   &     6.736~~     &  -2.685~~  \\
			4msn    &   -7.198~~    &    8.481~~     &  -0.849~~   \\
			5c1w     & -11.798~~     &   5.292~~     &  -2.229~~   \\
			5c28     & -10.404~~    &   15.595~~    &   -0.667~~   \\
			5c2h     & -22.453~~    &   -2.850~~    &    7.878~~   \\
			\hline 
			\hline
			PDB ID & $\Xi^\mathrm{pol}$~~ & $E^\mathrm{pair} + \Xi^\mathrm{pol}$ & $\Xi^\mathrm{pol}/(E^\mathrm{pair} + \Xi^\mathrm{pol})$ \\
			\hline
			1o3f  &  -18.158~~ &   27.304~~  &  -0.665~~  \\
			3ui7  & -16.902~~  & 51.054~~   & -0.331~~  \\
			3uuo &  -8.104~~ &  65.857~~  &  -0.123~~  \\
			\hline
			\hline
			PDB ID & $\Xi^\mathrm{pol}$~~ & $\Psi^\mathrm{bind,np} + \Xi^\mathrm{pol}$~~ & $\Xi^\mathrm{pol}/(\Psi^\mathrm{bind,np}+\Xi^\mathrm{pol})$~~ \\
			\hline
			2weg  &   -68.146~~   &   -53.759~~   &     1.268~~  \\
			3dx1 & -80.769~~ & -12.953~~ & 6.235~~  \\ 
			3dx2 & -73.340~~ & -29.526~~ & 2.484~~  \\ 
			3kwa   &   -77.169~~   &   -45.308~~    &    1.703~~   \\
			\hline
			\hline
		\end{tabular} 

\end{table}

\end{document}